\documentclass[sigconf]{acmart}
\usepackage{xcolor}
\AtBeginDocument{%
  }

\copyrightyear{2025} 
\acmYear{2025} 
\setcopyright{acmlicensed}
\acmConference[WWW '25] {Proceedings of the ACM Web Conference 2025}{April 28--May 2, 2025}{Sydney, NSW, Australia.}
\acmBooktitle{Proceedings of the ACM Web Conference 2025 (WWW '25), April 28--May 2, 2025, Sydney, NSW, Australia}
\acmISBN{979-8-4007-1274-6/25/04} 
\acmDOI{10.1145/3696410.3714677}
\usepackage{algorithm}
\usepackage{algorithmic}
\usepackage{amsthm}
\usepackage{booktabs} 
\usepackage{multirow} 
\usepackage{arydshln}
\usepackage{tikz}
\usepackage{xspace}
\usepackage{stmaryrd}
\usepackage{paralist}
\usepackage{mathtools}

\newcommand{\IndividualNames}{\mathcal{E}}
\newcommand{\RoleNames}{\mathcal{R}}
\newcommand{\ConceptNames}{\mathcal{C}}

\newcommand{\KBase}{\mathcal{K}}
\newcommand{\RBox}{\mathcal{R}}
\newcommand{\TBox}{\mathcal{T}}
\newcommand{\ABox}{\mathcal{A}}
\newcommand{\Int}{\mathcal{I}}

\newcommand{\RoleAssertion}[3]{#1(#2, #3)}
\newcommand{\ConceptAssertion}[2]{#1(#2)}

\newcommand{\nominal}[1]{\{#1\}}

\newcommand{\Individual}[1]{\mathsf{#1}}

\newcommand{\OscarsBestDirector}{\Individual{Oscar}}
\newcommand{\RelationName}[1]{\mathsf{#1}}
\newcommand{\wonBy}{\RelationName{wonBy}}
\newcommand{\edited}{\RelationName{edited}}
\newcommand{\produced}{\RelationName{produced}}

\newcommand{\SROIm}{\mathcal{SROI}^-}
\newcommand{\ALCOIR}{\mathcal{ALCOIR}}
\newcommand{\SROImR}{\operatorname{DAG}}

\newcommand{\TreeQuery}{\phi}
\newcommand{\TreeQueryOne}{\TreeQuery_1}
\newcommand{\TreeQueryTwo}{\TreeQuery_2}

\newcommand{\NonTreeQuery}{\psi}

\newcommand{\DLQuery}{C}
\newcommand{\DLQueryOne}{\DLQuery_1}
\newcommand{\DLQueryTwo}{\DLQuery_2}

\newcommand{\RDLQuery}{D}

\newcommand{\OurMethod}{DAGE}

\newcommand{\ComputationGraph}{\Gamma}
\newcommand{\ComputationNodes}{N}
\newcommand{\ComputationEdges}{E}
\newcommand{\ComputationNodeLabel}{\lambda}
\newcommand{\ComputationTarget}{\tau}
\newcommand{\Query}{Q}

\newcommand{\RoleComposition}[2]{{#1[#2]}}

\newcommand{\Approximated}{\operatorname{tree}}

\newcommand{\embedding}[1]{\mathbf{Emb}_{#1}}
\newcommand{\Nominal}{\operatorname{Nominal}}
\newcommand{\RelT}{\operatorname{RelT}}
\newcommand{\Intersection}{\operatorname{Intersect}}
\newcommand{\Complement}{\operatorname{Complement}}
\newcommand{\IndividualE}{\mathbf{E}}
\newcommand{\RoleE}{\mathbf{R}}
\newcommand{\QueryE}{\mathbf{Q}}
\newcommand{\Insideness}{\operatorname{Insideness}}
\newcommand{\Combiner}{\operatorname{RCombiner}}
\newcommand{\Composition}{\operatorname{RComposition}}
\newcommand{\Difference}{\operatorname{Diff}}
\newcommand{\Ce}[1]{\operatorname{Cen}(#1)}
\newcommand{\Of}[1]{\operatorname{Off}(#1)}
\newcommand{\Em}[1]{\mathbf{#1}}
\newcommand{\Loss}{\mathcal{L}}
\newcommand{\MLP}{\operatorname{MLP}}

\newcommand{\gb}[1]{\boldsymbol{#1}}
\newcommand{\taxis}{\text{ax}}
\newcommand{\targ}{\text{ap}}

\newcommand{\ee}{\mathbf{v}}
\newcommand{\qe}{\mathbf{V}_q}

\usepackage{color,soul}
\usepackage{subfigure}
\usepackage{orcidlink}
\begin{document}

\title{{\OurMethod}: DAG Query Answering via Relational Combinator with Logical Constraints}

\author{Yunjie He~\orcidlink{0009-0005-4461-2863}}
\authornote{Corresponding author}
\email{yunjie.he@ki.uni-stuttgart.de}
\affiliation{
  \institution{University of Stuttgart}
  \institution{Bosch Center for Artificial Intelligence}
  \city{Stuttgart}
  \country{Germany}
}

\author{Bo Xiong~\orcidlink{0000-0002-5859-1961}}
\email{xiongbo@stanford.edu}
\affiliation{
  \institution{Stanford University}
  \city{Palo Alto}
  \country{United States}
}

\author{Daniel Hernández~\orcidlink{0000-0002-7896-0875}}
\email{daniel.hernandez@ki.uni-stuttgart.de}
\affiliation{
  \institution{University of Stuttgart}
  \city{Stuttgart}
  \country{Germany}
}

\author{Yuqicheng Zhu~\orcidlink{0000-0001-5845-5401}}
\affiliation{
  \institution{University of Stuttgart}
  \institution{Bosch Center for Artificial Intelligence}
  \city{Stuttgart}
  \country{Germany}
}

\author{Evgeny Kharlamov~\orcidlink{0000-0003-3247-4166}}
\affiliation{
  \institution{Bosch Center for Artificial Intelligence}
  \city{Stuttgart}
  \country{Germany}
}

\author{Steffen Staab~\orcidlink{0000-0002-0780-4154}}
\affiliation{
  \institution{University of Southampton}
  \city{Southampton}
  \country{United Kingdom}
}
\affiliation{
  \institution{University of Stuttgart}
  \city{Stuttgart}
  \country{Germany}
}
\renewcommand{\shortauthors}{Yunjie He et al.}

\begin{abstract}
  Predicting answers to queries over knowledge graphs is called a complex reasoning task because answering a query requires subdividing it into subqueries. Existing query embedding methods use this decomposition to compute the embedding of a query as the combination of the embedding of the subqueries. This requirement limits the answerable queries to queries having a single free variable and being decomposable, which are called tree-form queries and correspond to the $\mathcal{SROI}^-$ description logic. In this paper, we define a more general set of queries, called DAG queries and formulated in the $\mathcal{ALCOIR}$ description logic, propose a query embedding method for them, called DAGE, and a new benchmark to evaluate query embeddings on them. Given the computational graph of a DAG query, DAGE combines the possibly multiple paths between two nodes into a single path with a trainable operator that represents the intersection of relations and learns DAG-DL concepts from tautologies. We implement DAGE on top of existing query embedding methods, and we empirically measure the improvement of our method over the results of vanilla methods evaluated in tree-form queries that approximate the DAG queries of our proposed benchmark.
\end{abstract}



\maketitle

\section{Introduction}%
\label{sec:Introduction}
A challenging aspect of machine learning is \emph{complex reasoning}, which involves solving tasks by breaking them into subtasks. A prominent complex reasoning problem is predicting answers to queries in knowledge graphs. This problem, called \emph{complex query answering}, involves solving queries by decomposing them into subqueries.  To tackle this, query embedding (QE) methods \citep{Query2Box, ConE, BetaE, BiQE} encode queries as low-dimensional vectors and use neural logical operators to combine subquery embeddings into a single query embedding. However, these methods are limited to processing first-order logic queries with a single unquantified target variable, known as \emph{tree-form queries}, which align with $\SROIm$ description logic~\citep{AConE} and have tree-like computation graphs~\cite{CLQA_Survey}. In contrast, this work explores \emph{DAG queries}, a more expressive set that extends tree-form queries by allowing quantified variables to appear multiple times in atom components. This enables multiple paths from a quantified variable $x$ to a target variable $y$, unlike tree-form queries, which permit at most one such path. 

Consider the following first-order query $\TreeQuery(x)$, asking for works edited by an Oscar winner and produced by an Oscar winner.
\begin{equation}
\TreeQuery(y) \Coloneqq \exists x_1 \exists x_2 :
  \begin{aligned}[t]
    & \wonBy(\OscarsBestDirector, x_1)\; \land \edited(x_1, y)\; \land \\[-3pt]
    & \wonBy(\OscarsBestDirector, x_2)\; \land \produced(x_2, y). \\[-3pt]
  \end{aligned}
  \label{eq:tree-form}
\end{equation}

The computation graph of query $\TreeQuery(y)$ is the following:
\begin{equation}
  \begin{aligned}[c]
  \begin{tikzpicture}
  \node(a) at (0,0) {$\OscarsBestDirector$};
  \node(x1) at (3.1,0.3) {$x_1$};
  \node(x2) at (3.1,-0.3) {$x_2$};
  \node(y) at (6,0) {$y$};
  \draw[->] (a) edge[bend left] node[above] {$\wonBy$} (x1);
  \draw[->] (a) edge[bend right] node[below] {$\wonBy$} (x2);
  \draw[->] (x1) edge[bend left] node[above] {$\edited$} (y);
  \draw[->] (x2) edge[bend right] node[below] {$\produced$} (y);
  \end{tikzpicture}
  \end{aligned}
\end{equation}
Query $\TreeQuery(y)$ is tree-form because there exists at most one path from $x_1$ to $y$ and at most one path from $x_2$ to $y$. Since it is tree-form, it can be expressed as a conjunction $\TreeQueryOne(y) \land \TreeQueryTwo(y)$, where the subqueries $\TreeQueryOne(y)$ and $\TreeQueryTwo(y)$ are also tree-form:
\begin{align}
  \TreeQueryOne(y)
  &\Coloneqq \exists x_1 : 
    \wonBy(\OscarsBestDirector, x_1) \land \edited(x_1, y), \\
  \TreeQueryTwo(y)
  &\Coloneqq \exists x_2 :
    \wonBy(\OscarsBestDirector, x_2)\; \land
    \produced(x_2, y).
\end{align}
In $\SROIm$, query $\TreeQuery$ is expressed as $\DLQuery = \DLQueryOne \sqcap \DLQueryTwo$, where the subqueries $\DLQueryOne$ and $\DLQueryTwo$ are:
\begin{align}
  \DLQueryOne
  &\Coloneqq \exists \edited^-.(\exists \wonBy^-.\nominal{\OscarsBestDirector}), \TreeQueryTwo(y)
  \label{eq:example-concept1} \\
  \DLQueryTwo
  &\Coloneqq \exists \produced^-.(\exists \wonBy^-.\nominal{\OscarsBestDirector}).
  \label{eq:example-concept2}
\end{align}
Complex query answering methods use the embeddings of the subqueries $\TreeQueryOne$ and $\TreeQueryTwo$ to compute the embedding of query $\TreeQuery$. To this end, these methods define a neural logical operator that represents the logical operation $\land$. Hence, the ability to decompose queries into subqueries and express the logical connectives with neural logical operators is critical for the existing complex reasoning methods.

We show that a query where this decomposition of queries does no longer hold. Consider the query asking for works edited and produced by an Oscar winner (i.e., an Oscar winner that has both roles, editor and producer). Compared with the previous query, this new query enforces $x_1 = x_2$, which can be encoded by renaming both variables $x_1$ and $x_2$ as $x$:
\begin{equation}
\NonTreeQuery(y) = \exists x :
  \begin{aligned}[t]
    & \wonBy(\OscarsBestDirector, x)\; \land \\[-3pt]
    & \edited(x, y)\; \land \produced(x, y). \\[-3pt]
  \end{aligned}
  \label{eq:non-tree-form}
\end{equation}
If we observe the computation graph of query $\NonTreeQuery(y)$, depicted in \eqref{eq:DAG-fo-computation-graph}, we can see that $\NonTreeQuery(y)$ is no longer a tree-form query because there are two paths from variable $x$ to variable $y$. We call Directed Graph Queries (DAG) to the queries in the set resulting from relaxing the maximum of one path restriction of tree-form queries.
\begin{equation}
  \begin{aligned}[c]
  \begin{tikzpicture}
  \node(a) at (0,0) {$\OscarsBestDirector$};
  \node(x) at (3,0) {$x$};
  \node(y) at (6,0) {$y$};
  \draw[->] (a) edge[edge label=$\wonBy$] (x);
  \draw[->] (x) edge[bend left, edge label=$\edited$] (y);
  \draw[->] (x) edge[bend right] node[below] {$\produced$} (y);
  \end{tikzpicture}
  \end{aligned}
  \label{eq:DAG-fo-computation-graph}
\end{equation}
Query $\NonTreeQuery(y)$ cannot be decomposed into two tree-form queries because the conjunction between the query atoms $\edited(x, y)$ and $\produced(x, y)$ requires considering two target variables in the complex reasoning subtask.
Similarly, query $\NonTreeQuery(y)$ is not expressible in $\SROIm$ because $\SROIm$ allows for conjunctions in concept descriptions but not role descriptions, which are required to indicate that $x$ ``$\produced$ and $\edited$'' $y$.
A description logic that allows for conjunctions in role descriptions, called $\ALCOIR$\footnote{$\ALCOIR$ is a description logic in the family of Attributive Languages ($\mathcal{AL}$). The letters $\mathcal{C}$, $\mathcal{O}$, $\mathcal{I}$, and $\mathcal{R}$ stand for the extensions to the $\mathcal{AL}$ description logic with complement, nominals, inverse predicates, and conjunction of role descriptions~\cite{DBLP:conf/dlog/2003handbook}.}, can express the query $\NonTreeQuery(y)$ as the following concept description $D$\footnote{In this paper we use the terms concept description and query as synonyms because a concept description defines the answers to the query.}:
\begin{equation}
  \RDLQuery \Coloneqq
  \exists (\edited \sqcap \produced)^-.(\exists \wonBy^-.\nominal{\OscarsBestDirector}).
  \label{eq:dag-query-d}
\end{equation}

Unlike existing methods, to compute the embedding of query $D$, we do not decompose $D$ into two subqueries, but we compute the embedding of the relation description $\edited \sqcap \produced$ with an additional neural operator, called \emph{relational combinator}, to represent the intersection between relations. With this extension to existing methods~\citep{Query2Box,BetaE, ConE}, we can represent the aforementioned query $D$ with the following computation graph.
\begin{equation}
  \begin{aligned}[c]
  \begin{tikzpicture}
  \node(a) at (0,0) {$\OscarsBestDirector$};
  \node(x) at (2.6,0) {$x$};
  \node(y) at (6,0) {$y$};
  \draw[->] (a) edge[edge label=$\wonBy$] (x);
  \draw[->] (x) edge[edge label=$\edited \sqcap \produced$] (y);
  \end{tikzpicture}
  \end{aligned}
\end{equation}
Like the computation graphs of tree-form queries, the computation graph of query $D$ has a single path from variable $x$ to variable $y$. Hence, we can reuse existing query embedding methods~\citep{Query2Box,BetaE, ConE} to recursively define the embedding of a DAG query.

Without our extension, the vanilla methods cannot be applied to DAG queries. Instead, we can apply them by relaxing DAG queries to tree-form queries. However, one can expect that this workaround solution produces less accurate results because the solutions to the relaxed query $\TreeQuery(y)$ in \eqref{eq:tree-form} are not enforced to satisfy $x_1 = x_2$ like the solutions to query $\NonTreeQuery(y)$ in \eqref{eq:non-tree-form}.

To define the relational combinator for role conjunctions, we encourage the models to satisfy a set of well-known $\ALCOIR$ tautologies involving role conjunctions.

\medskip
\noindent
In summary, this paper makes the following contributions:
\begin{enumerate}
\item In Section~\ref{sec:DAG-DL}, we propose a description logic, named $\SROImR$, that extends $\SROIm$ to encode conjunction of relations, and we present four tautologies involving this extension.
\item In Section~\ref{sec:dag-query-answering}, we propose an integrable relational combinator that can be integrated into existing query embedding methods and generally enhance their expressiveness to $\SROImR$ queries, and that follows three tautologies involving the intersection of relations.
\item In Section \ref{Query Generation}, we introduce six novel types of $\SROImR$ queries, and their corresponding relaxed tree-form queries, and develop new datasets with different test difficulty levels.
\item In Section \ref{Main Results}, we assess the performance of existing methods on the created datasets, comparing them with our integrated module. The results show that {\OurMethod} brings significant improvement to the baseline models on DAG queries. 
\item In Section \ref{section:ablation study}, we create new data splits on the benchmark datasets to analyze {\OurMethod}'s effectiveness in improving query embedding models for DAG queries in greater detail. Our results show significant improvements over exiting approaches. 
\end{enumerate}


\section{Preliminaries}%
\label{sec:preliminaries}


This section presents queries as $\ALCOIR$ concepts. We follow the notations and semantics described in~\cite{DBLP:conf/dlog/2003handbook}.
For the following definitions, we assume three pairwise disjoint sets $\ConceptNames$, $\RoleNames$, and $\IndividualNames$, whose elements are called \emph{concept names}, \emph{role names} and \emph{individual names}.

\begin{definition}[Syntax of $\ALCOIR$ Concept and Role Descriptions]%
  \label{def:ACLR-queries}
  $\ALCOIR$ \emph{concept descriptions} $C$ and \emph{role descriptions} $R$ are defined by the following grammar
  \begin{align*}
     C &\Coloneqq
      \top \mid A \mid \nominal{a} \mid \neg C \mid C \sqcap C
      \mid \exists R.C \\
     R &\Coloneqq r \mid R^- \mid R \circ R \mid R \sqcap R \mid R^+
  \end{align*}
  where the symbol $\top$ is a special concept description, and symbols $A$, $a$ and $r$ stand for concept names, individual names, and role names, respectively. Concept descriptions $\nominal{a}$ are called \emph{nominals}.
  We write $\bot$, $C \sqcup D$, $\forall R.C$ as abbreviations for $\neg\top$, $\neg(\neg C \sqcap \neg D)$ and $\neg\exists R.\neg C$, respectively.
\end{definition}

\begin{definition}[Syntax of $\ALCOIR$ Knowledge Bases]
  Given two $\ALCOIR$ concept descriptions $C$ and $D$ and two role descriptions $R$ and $S$, the expressions $C \sqsubseteq D$ and $R \sqsubseteq S$ are respectively \emph{concept inclusion} and a \emph{role inclusion}. We write $C \equiv D$ as an abbreviation for two concept inclusions $C \sqsubseteq D$ and $D \sqsubseteq C$, and likewise for $R \equiv S$. Given two individual names $a$ and $b$, a concept description $C$ and a relation description $R$, the expression $\ConceptAssertion{C}{a}$ is a \emph{concept assertion} and the expression $\RoleAssertion{R}{a}{b}$ is a \emph{role assertion}. An $\ALCOIR$ knowledge base is a triple $\KBase = (\RBox, \TBox, \ABox)$ where $\RBox$ is a set of role inclusions, $\TBox$ is a set of concept inclusion, and $\ABox$ is a set of concept and role assertions.
\end{definition}

\begin{definition}[Interpretations]
  An \emph{interpretation} $\Int$ is a tuple $(\Delta^\Int, \cdot^\Int)$ where $\Delta^\Int$ is a set and $\cdot^\Int$ is a function with domain $\IndividualNames \cup \ConceptNames \cup \RoleNames$, called the \emph{interpretation function}, that maps every individual name $a \in \IndividualNames$ to an element $a^\Int \in \Delta^\Int$, every concept name $A \in \ConceptNames$ to a set $A^\Int \subseteq \Delta^\Int$, and every role name $r \in \RoleNames$ to a relation $r^\Int \subseteq \Delta^\Int \times \Delta^\Int$.
  The interpretation function is recursively extended to $\ALCOIR$ concept descriptions and role descriptions by defining the semantics of each connective (see~\cite{DBLP:conf/dlog/2003handbook} and Appendix~\ref{sec:ACOIR-descriptions}).
\end{definition}

\begin{definition}[Semantics of $\ALCOIR$ Knowledge Bases]
  Given an interpretation $\Int$, we say that $\Int$ is a \emph{model} of
  \begin{itemize}
  \item a role axiom $R \sqsubseteq S$ if and only if $R^\Int \subseteq S^\Int$,
  \item a concept axiom $C \sqsubseteq D$ if and only if $C^\Int \subseteq D^\Int$,
  \item a concept assertion $\ConceptAssertion{C}{a}$ if and only if $a^\Int \in C^\Int$,
  \item a role assertion $\RoleAssertion{R}{a}{b}$ if and only if $(a^\Int, b^\Int) \in R^\Int$,
  \item an $\ALCOIR$ knowledge base $\KBase = (\RBox, \TBox, \ABox)$ if and only if $\Int$ is a model of every element in $\RBox \cup \TBox \cup \ABox$.
  \end{itemize}
\end{definition}

\begin{definition}[Entailment]
  Given two knowledge bases $\KBase_1$ and $\KBase_2$ we say that $\KBase_1$ \emph{entails} $\KBase_2$, denoted $\KBase_1 \models \KBase_2$, if for every interpretation $\Int$, if $\Int$ models $\KBase_1$ then $\Int$ models $\KBase_2$. This definition is extended to axioms and assertions (e.g., $\KBase \models \ConceptAssertion{C}{a}$ if all models or $\KBase$ are also models of $\ConceptAssertion{C}{a}$).
\end{definition}

\begin{definition}[Knowledge Graph~\cite{AConE}]%
  \label{def:kg}
  A knowledge graph $G$ is an $\ALCOIR$ knowledge base whose RBox is empty, its TBox contains a unique concept inclussion $\top \sqsubseteq \{a_1\} \sqcup \cdots \sqcup \{a_n\}$, called \emph{domain-closure assumption}, where $\{a_1,\dots,a_n\}$ is the set of all individuals names occurring in the ABox, and its ABox contains only role assertions.
\end{definition}

\begin{definition}[Knowledge Graphs Query Answers]
  Given a knowledge graph $G$, the answers to an $\ALCOIR$ concept description $C$ are the individual names $a \in \IndividualNames$ such that $G \models C(a)$.
\end{definition}


\section{Tree-Form and the DAG Queries}
\label{sec:DAG-DL}

As was proposed by He et al.~\cite{AConE}, tree-form queries can be expressed as $\SROIm$ concepts descriptions. The computation graphs of the first-order formulas corresponding to these concepts descriptions have at most one path for every quantified variable to the target variable. As we already show, this does not hold if the relational intersection $\sqcap$ is added. In this section, we define tree-form and DAG queries as subsets of the $\ALCOIR$ description logic, we describe their computation graph, and the relaxation of non-tree form DAG queries as tree-form queries.

\subsection{Syntax of Queries}

\begin{definition}[Tree-Form and DAG queries]%
  \label{def:tree-form-queries}
  \emph{DAG queries} are the subset of $\ALCOIR$ concept descriptions $C$ defined by the following grammar
  \begin{align*}
     C &\Coloneqq
      \nominal{a} \mid \neg C \mid C \sqcap C
      \mid \exists R.C \\
     R &\Coloneqq r \mid R^- \mid R \circ R \mid R \sqcap R
  \end{align*}
  A DAG query is said tree-form if it does not include the operator $\sqcap$ in role descriptions.
\end{definition}

Unlike $\ALCOIR$ concept descriptions, DAG queries do not include concept names, the $\top$ concept, nor the transitive closure or relations. We exclude these constructors because they are not present in queries supported by existing query embeddings. 

\begin{proposition}%
  \label{prop:tautologies}
  Given two role descriptions $R$ and $S$, and an individual name $a$, the following equivalences hold:
  \begin{tabbing}
  commutativity: \hspace{7em}\=  $R \sqcap S \equiv S \sqcap R$,\\
  monotonicity: \> $R \sqcap S \sqsubseteq R$,\\
  restricted conjunction preserving: \> $\exists (R \sqcap S). \nominal{a} \equiv \exists R.\nominal{a} \sqcap \exists S.\nominal{a}$.
  \end{tabbing}
\end{proposition}

\begin{proof}
  The tautologies follow directly from the semantics of $\ALCOIR$ concept and role descriptions.
\end{proof}

\subsection{Computation Graphs}%
\label{sec:computation-graphs}

He et al.~\cite{AConE} illustrated the computation graphs for tree-form queries encoded as $\SROIm$ concepts, but did not formalize them. We next provide such a formalization for a graph representation of DAG queries (and thus for tree-form queries).

\begin{definition}[Computation Graph]
  A \emph{computation graph} is a labelled directed graph $\ComputationGraph = (\ComputationNodes, \ComputationEdges,\ComputationNodeLabel, \ComputationTarget)$ such that $\ComputationNodes$ is a set whose elements are called \emph{nodes}, $\ComputationEdges \subseteq \ComputationNodes \times \ComputationNodes$ is a set whose elements are called \emph{edges}, $\ComputationNodeLabel$ is a function that maps each node in $\ComputationNodes$ to a \emph{label}, and $\ComputationTarget$ is a distinguished node in $\ComputationNodes$, called \emph{target}.
\end{definition}

\begin{example}%
  \label{ex:base-computation-graph}
  The computation graph $\ComputationGraph = (\ComputationNodes, \ComputationEdges, \ComputationNodeLabel, \ComputationTarget)$ with $\ComputationNodes = \{u_1, u_1\}$, $\ComputationEdges = \{(u_1, u_2)\}$, $\ComputationNodeLabel = \{u_1 \mapsto \nominal{\OscarsBestDirector}, u_2 \mapsto \exists\wonBy^-\}$, and $\ComputationTarget = u_2$ is depicted in \eqref{eq:computation-graph-base}.
    \begin{equation}
    \begin{aligned}[c]
      \begin{tikzpicture}[%
          grow=180,%
          level distance=2.1cm,%
          sibling distance=0.6cm,%
          nodes={draw, rounded corners, font=\scriptsize, minimum height=1.3em},%
          edges from parent/.style={draw}
          ]
          \node {$u_2 : \exists\wonBy^-$}
          child {
            node {$u_1 : \nominal{\OscarsBestDirector}$}
          };
      \end{tikzpicture}
    \end{aligned}
    \label{eq:computation-graph-base}
  \end{equation}
  Intuitively, the node $u_1$ computes the concept $\nominal{\OscarsBestDirector}$ and the node $u_2$ computes the concept $\exists \wonBy^-.\nominal{\OscarsBestDirector}$, which corresponds to answers to the query asking who is an Oscar's winner.
\end{example}

To define the computation graphs of DAG queries, we need to introduce the composition of a computation graph $\ComputationGraph$ with a role description $R$, denoted $\RoleComposition{\ComputationGraph}{R}$. Intuitively, the composition is the concatenation of $\ComputationGraph$ with the graph representing the role description, as Example~\ref{ex:role-composition} illustrates. A definition for this operation and the computation graph of DAG queries is suplemented in Appendix~\ref{sec:computation-graphs-appendix}.

\begin{example}
  \label{ex:role-composition}
  Consider the computation graph $\ComputationGraph$ depicted in \eqref{eq:computation-graph-base}. The computation graph $\RoleComposition{\ComputationGraph}{\edited^-}$ and $\RoleComposition{\ComputationGraph}{(\edited \sqcap \produced)^-}$ are depicted in \eqref{eq:computation-graph-extended-1} and \eqref{eq:computation-graph-extended-2}, respectively.
  \begin{equation}
    \begin{aligned}[c]
      \begin{tikzpicture}[%
          grow=180,%
          level distance=2.1cm,%
          sibling distance=0.6cm,%
          nodes={draw, rounded corners, font=\scriptsize, minimum height=1.3em},%
          edges from parent/.style={draw}
          ]
          \node {$u_3 : \exists\edited^-$}
          child {
            node {$u_2 : \exists\wonBy^-$}
            child {
              node {$u_1 : \nominal{\OscarsBestDirector}$}
            }
          };
      \end{tikzpicture}
    \end{aligned}
    \label{eq:computation-graph-extended-1}
  \end{equation}
  \begin{equation}
    \begin{aligned}[c]
      \begin{tikzpicture}[%
          grow=180,%
          level distance=2.1cm,%
          sibling distance=0.6cm,%
          nodes={draw, rounded corners, font=\scriptsize, minimum height=1.3em},%
          edges from parent/.style={draw}
          ]
          \node at (0,0) {$u_5 : \sqcap$}
          child {
            node(a) {$u_3 : \exists\edited^-$}
          }
          child {
            node(b) {$u_4 : \exists\produced^-$}
          };
          \node(c) at (-4.3,0) {$u_2 : \exists\wonBy^-$}
          child {
            node {$u_1 : \nominal{\OscarsBestDirector}$}
          };
          \path[draw] (c)--(a);
          \path[draw] (c)--(b);
      \end{tikzpicture}
    \end{aligned}
    \label{eq:computation-graph-extended-2}
  \end{equation}
\end{example}

\begin{example}
  Consider the tree-form query $C = C_1 \sqcap C_2$ defined by equations \eqref{eq:example-concept1} and \eqref{eq:example-concept2}. The computation graph of $C$ is depicted in the following diagram:
  \begin{equation}
    \begin{aligned}[c]
      \begin{tikzpicture}[%
          grow=180,%
          level distance=2.1cm,%
          sibling distance=0.6cm,%
          nodes={draw, rounded corners, font=\scriptsize, minimum height=1.3em},%
          edges from parent/.style={draw}
          ]
          \node  {$u_4 : \sqcap$}
          child {
            node {$u_3 : \exists\edited^-$}
            child {
              node {$u_2 : \exists\wonBy^-$}
              child {
                node {$u_1 : \nominal{\OscarsBestDirector}$}
              }
            }
          }
          child {
            node {$u_7 : \exists\produced^-$}
            child {
              node {$u_6 : \exists\wonBy^-$}
              child {
                node {$u_5 : \nominal{\OscarsBestDirector}$}
              }
            }
          };
      \end{tikzpicture}
    \end{aligned}
    \label{eq:tree-form-computation-graph}
  \end{equation}
  Similarly, the computation graph for the DAG query in equation \eqref{eq:dag-query-d} is depicted by the figure in \eqref{eq:computation-graph-extended-2}.
  In \eqref{eq:tree-form-computation-graph}, different nodes can have the same label and represent the same concept (e.g., nodes $u_2$ and $u_6$ represent the concept $\exists\wonBy^-.\nominal{\OscarsBestDirector}$). Intuitively, the duplication of labels means that an answer can be a work edited by an Oscar's winner and produced by another Oscar's winner. On the other hand, since there is a single node for this concept in the computation graph in \eqref{eq:computation-graph-extended-2}, namely node $u_2$, the work must be produced and edited by the same Oscar's winner.
\end{example}

\subsection{Relaxing Non Tree-Form DAG queries}

The restricted conjunction preserving (see Proposition~\ref{prop:tautologies}), does no longer follow if we replace the nominal $\nominal{a}$ with a general concept description $C$. Indeed, the example discussed in the introduction is a counterexample for the generalized version of the conjunction preserving. The fact that conjunction is not preserved in general is the cause of the need of new neural operator for the role conjunction, different from the one used for the concept conjunction. Alternatively, if the neural operator is not used, we can relax concept description with a relaxed version of this tautology.

\begin{proposition}
  Given two role descriptions $R$ and $S$, and a concept description $C$, the following concept inclusion holds:
  \begin{equation}
      \exists(R \sqcap S).C \sqsubseteq \exists R.C \sqcap S.C
      \label{eq:relaxing}
  \end{equation}
\end{proposition}

\begin{proof}
    By monotonicity, concept $\exists(R \sqcap S).C$ is included in the concepts $\exists R.C$ and $\exists S.C$. Then, concept $\exists(R \sqcap S).C$ is included in the concept $\exists R.C \sqcap \exists S.C$.
\end{proof}

Intuitively, the role conjunction relaxation consists of not assuming that the instances of the concept $C$ must be equal on the concept defined on the right side. An example of this was discussed in the introduction, when the editor and producer of a work are not required to be the same Oscar's winner. Thus, the three-form query with the computation graph in \eqref{eq:tree-form-computation-graph} relaxes the non tree-form with the computation graph in \eqref{eq:computation-graph-extended-2}.

\begin{definition}[Tree-form approximation]
    The \emph{approximated tree-form query} of a DAG query $Q$, denoted $\Approximated(Q)$, is the tree-form query resulting from removing every conjunction of role descriptions using the inclusion in \eqref{eq:relaxing}.
\end{definition}

It is not difficult to see that for every DAG query $Q$ with no complement constructor (i.e., without $\neg$), it holds that $Q \sqcap \Approximated(Q)$, that is, query $\Approximated(Q)$ relaxes query $Q$. This is not necessary for queries including complement because they are not necessarily monotonic. Hence, query embeddings that use tree-form queries to predict answers to DAG queries are expected to incur in both, more false positives and more false negatives.



\section{DAG Query Answering with Relational Combinator}
\label{sec:dag-query-answering}

In this section, we first introduce a generalized query embedding model subsuming various previous query embedding approaches \citep{Query2Box,BetaE,ConE}. 
Then, we introduce a relational combinator that extends existing query embeddings to support DAG query type. 
Finally, we discuss how to introduce additional logical constraints to further improve the results. 

\subsection{Base Query Embedding Methods Interface}
Many query embedding methods~\citep{Query2Box,BetaE,ConE} predict query answers by comparing the embedding of individuals with the embedding of the query, so the individuals that are closer to the query in the embedding space are more likely to be answers. These query embeddings are learnable parameterized objects and are computed via neural operations that correspond to the logic connectives in the queries. In this subsection we present the interface required for the embedding methods to be used as a base for our proposed query embedding method, DAGE. Query embedding methods such as Query2Box, BetaE, and ConE satisfy this interface. 

We assume three vector spaces $\IndividualE^d$, $\RoleE^d$, and $\QueryE^d$, where $\IndividualE$, $\RoleE$, and $\QueryE$ are fields (which depend on the query embedding method) and $d$ is the dimension of the vectors.
We assume that every individual $a$ is embedded in a vector $\embedding{a} \in \IndividualE^d$, every role name $r$ and its inverse $r^-$ are embedded in vector $\embedding{r}, \embedding{r^-} \in \RoleE^d$, and every tree-form query $\Query$ is embedded in a vector $\embedding{\Query} \in \QueryE^d$. Whereas the embedding function $\embedding{\cdot}$ is defined for individuals and role names and the inverse of role names, because they are directly defined by the parameters to be learn, function $\embedding{\cdot}$ is not directly defined for compound role and concept descriptions.

\subsubsection{Role Embeddings}
The embedding of a role description $R$ is recursively computed from the embedding of role names and its inverses as with a neural operators with signature $\Composition : \RoleE^d \times \RoleE^d \to \RoleE^d$ as follows:
\begin{align}
  \embedding{R \circ S} &\Coloneqq \Composition(\embedding{R}, \embedding{S}), \\
  \embedding{R^{--}} &\Coloneqq \embedding{R}, \\
  \embedding{(R \circ S)^-} &\Coloneqq \embedding{S^- \circ R^-}.
\end{align}

\subsubsection{Concept  Embeddings}
The embedding of a query is computed from the embedding of individual names and role names using neural operators that represent the logical connectives in queries. The signatures of these neural operators are the following: $\Nominal: \IndividualE^d \to \QueryE^d$, $\RelT: \QueryE^D \times \RoleE^E \to \QueryE^E$, $\Intersection: \QueryE^d \times \ldots \QueryE^d \to \QueryE^d$, and $\Complement: \QueryE^d \to \QueryE^d$.
These neural operators define query embedding of tree-form queries as follows:
\begin{align}
  \embedding{\nominal{a}} &\Coloneqq \Nominal(\embedding{a}), \\
  \embedding{\exists r.C} &\Coloneqq \RelT_r(\embedding{C}), \\
  \embedding{C_1 \sqcap C_2 \ldots \sqcap C_n} &\Coloneqq \Intersection(\embedding{C_1}, \embedding{C_2}, \cdots,\embedding{C_n}),\\
  \embedding{\neg C} &\Coloneqq \Complement(\embedding{C}).
\end{align}

\subsubsection{Insideness}
Given a query $Q$ and a knowledge graph $G$, the goal of query embedding approaches is to maximize the predictions of \emph{positive answers} to query $\Query$ (i.e., individuals $a$ such that $G \models \ConceptAssertion{\Query}{a}$) and minimize the prediction of \emph{negative answers} to query $\Query$ (i.e., individuals $b$ such that $G \models \neg\ConceptAssertion{\Query}{b}$). Because of the open-world semantics of $G$ (see Definition~\ref{def:kg}) we cannot know which answers are negative. However, the learning of query embedding needs negative answers. Therefore, for each positive answer $a$, query embedding methods assume a random individual $b$, different from $a$, to be a negative answer.

In the representation space, the evaluation of how likely an individual is an answer to a query is computed with a function with signature $\Insideness: \QueryE^d \times \QueryE^d \to \mathbb{R}$, that returns higher numbers for individuals that are answers to the query than for individuals that are not answers to the query. That is, given a query $Q$ with a positive answer $a$ and its corresponding randomly generated negative answer $a'$ distinct from $a$, the goal of query embedding approaches is to minimize the following loss:
\begin{equation}
  \Loss_i(Q) \Coloneqq \sum_{a \in \IndividualNames} \left(
  \begin{aligned}[c]
    & - \log \sigma \left(\gamma - \Insideness(\embedding{\Query}, \embedding{\nominal{a}}) \right) \\
    & + \sum_j^k \frac{1}{k} \log \sigma \left( \gamma - \Insideness(\embedding{\Query}, \embedding{\nominal{a'}}) \right)
  \end{aligned}
  \right).
\end{equation}
where $\nominal{a'} $ us the negative sample, $\gamma$ is a margin hyperparameter and $k$ is the number of random negative samples for each positive query answer pair.




\subsection{The Relational Combinator}

So far, we have described an interface consisting of neural operators that are implemented by existing query embeddings. These neural operators allow the computation of tree-form query embeddings, but not DAG queries including the conjunction of roles. To enhance the capability of these methods for DAG queries, we introduce a relational combination operator with signature
\begin{equation}
  \Combiner_k : (\RoleE^d )^k \to \RoleE^d,
\end{equation}
where $k$ is a positive natural number. The embedding of a role description $R_1 \sqcap \cdots \sqcap R_k$ (with $k > 0$) is:
\begin{equation}
  \embedding{ R_1 \sqcap \cdots \sqcap R_k }
  \Coloneqq \Combiner_k( \embedding{R_1}, \dots, \embedding{R_k} ),
\end{equation}
where $\Combiner$ is a commutative neural network. We used the neural operator DeepSet~\cite{deepsets} to implement $\Combiner$.
\begin{equation}
  \Combiner_k(\embedding{R_1}, \dots, \embedding{R_k} ) = \sum_{1 \leq i \leq k} \alpha _{i} \cdot \text{MLP}(\embedding{R_i} ))
  \label{eq:deepset}
\end{equation}
where the weights $\alpha_1, \cdots, \alpha_k$ sum $1$. Specifically,
\[\alpha_{i} = \frac{\exp(\MLP(\embedding{R_i}))}{\sum_{1 \leq j \leq k}\exp(\MLP(\embedding{R_j}))}.\]


\begin{proposition}%
  \label{prop:tautologies-1}
  Given two role descriptions $R$ and $S$, 
  \begin{equation}
    \Combiner_2(\embedding{R}, \embedding{S}) = \Combiner_2(\embedding{S}, \embedding{R}).
  \end{equation}
\end{proposition}

\begin{proof}
  It follows from the commutativity of the DeepSet.
\end{proof}

Proposition~\ref{prop:tautologies-1} guarantees that the embedding of role description satisfies some of the $\ALCOIR$ tautologies described in Proposition~\ref{prop:tautologies}, namely commutativity and idempotence.

\subsection{Encouraging Tautologies}

So far, we have shown (see Proposition~\ref{prop:tautologies-1}) that the proposed relational combinator satisfies two of the $\ALCOIR$ tautologies presented in Proposition~\ref{prop:tautologies}, namely commutative and idempotence, but not the monotonicity and the restricted conjunction preserving. We hypothesize that by encouraging the query embeddings such that the inference of embeddings follow these tautologies, we can improve the embedding generalization capacity.




\subsubsection{Monotonicity}  We encourage the query embeddings of a DAG query $Q = \exists (R \sqcap S).C$ to be subsumed by the query embedding of query $Q' = \exists R.C$ by introducing the following loss:
\begin{equation}
  \Loss_m(Q) = \sum_{r \in \mathcal{R}, s \in \mathcal{R} } \Insideness( \embedding{Q}, \embedding{Q'} ),
\end{equation}
$\Insideness$ measures the likelihood of  $\embedding{Q'}$ being inside $\embedding{Q}$.\footnote{More details about the $\Insideness$ function for individual methods can be found in Appendix \ref{computations of baseline models}.} 

\subsubsection{Restricted conjunction preserving}
We encourage the tautology $\exists (r \sqcap s). \nominal{e} \equiv \exists r.\nominal{e} \sqcap \exists s.\nominal{e}$ (see Proposition~\ref{prop:tautologies}) with the following loss:
\begin{equation}
    \mathcal{L}_r \Coloneqq \Difference(\embedding{\exists (r \sqcap s). \nominal{e}}, \Intersection(\embedding{r.\nominal{e}}, \embedding{s.\nominal{e}})),
\end{equation}
where $\Difference$ measures the distance between two query embeddings. We supplement the details on $\Difference$ of each individual query embedding method in Appendix \ref{computations of baseline models}.

By imposing these loss terms, the tautologies are encoded into geometric constraints, which are soft constraints over the embedding space. Hence, our loss terms can also be viewed as regulations that reduce the embedding search space.
Given a query answering training dataset $\mathcal{D}$, our final optimization objective is:
\begin{equation}
    \Loss(D) \Coloneqq \sum_{i=1}^{|\mathcal{D}|} \Loss_i(Q) + \lambda_1 \Loss_m + \lambda_2 \mathcal{L}_r,
\end{equation}
where $\mathcal{L}_q$ is the query embedding loss, and $\lambda_1$ and $\lambda_2$ are the weights of regularization terms. 



\section{Experiments}
In this section, we answer the following research questions with experimental statistics and corresponding case analyses. \textbf{RQ1:} How effective is {\OurMethod} for enhancing baseline models on DAG queries? \textbf{RQ2:} How well does {\OurMethod} perform on tree-form queries? \textbf{RQ3:} How do the logical constraints influence the performance of {\OurMethod}? Details on experimental setups can be found in Appendix \ref{Experimental setup}.

\begin{figure}[t]
    \centering
    \includegraphics[width=\linewidth]{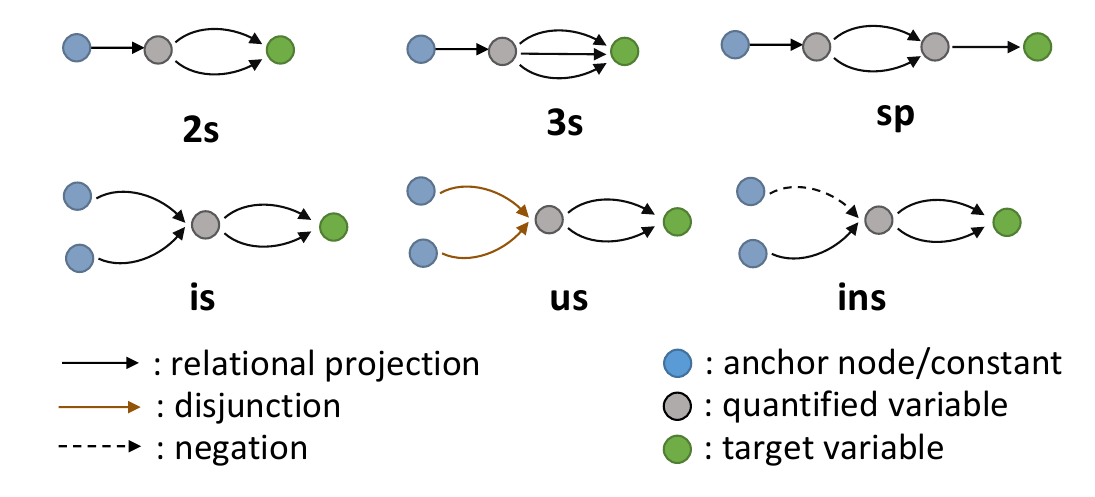}
    \caption{Query structures considered in the experiments, where anchor entities and relations are to be specified to instantiate logical queries. Each structure is named using abbreviations: "s" for split, "p" for projection, "i" for intersection, "u" for union, and "n" for negation. For instance, "2s" indicates a structure with two splitting edges.}
    \label{fig:DAG query types}
\end{figure}

\subsection{DAG Query Generation}
\label{Query Generation}
Existing datasets, e.g. NELL-QA, FB15k237-QA, WN18RR-QA, do not contain DAG queries. 
We propose six new DAG query types, i.e., 2s, 3s, sp, us, is, and ins, as shown in Figure \ref{fig:DAG query types}. Following these new query structures, we generate new DAG query benchmark datasets, NELL-DAG, FB15k-237-DAG, and FB15k-DAG.

\begin{figure}[t!]
    \centering
    \includegraphics[width=0.8\linewidth]{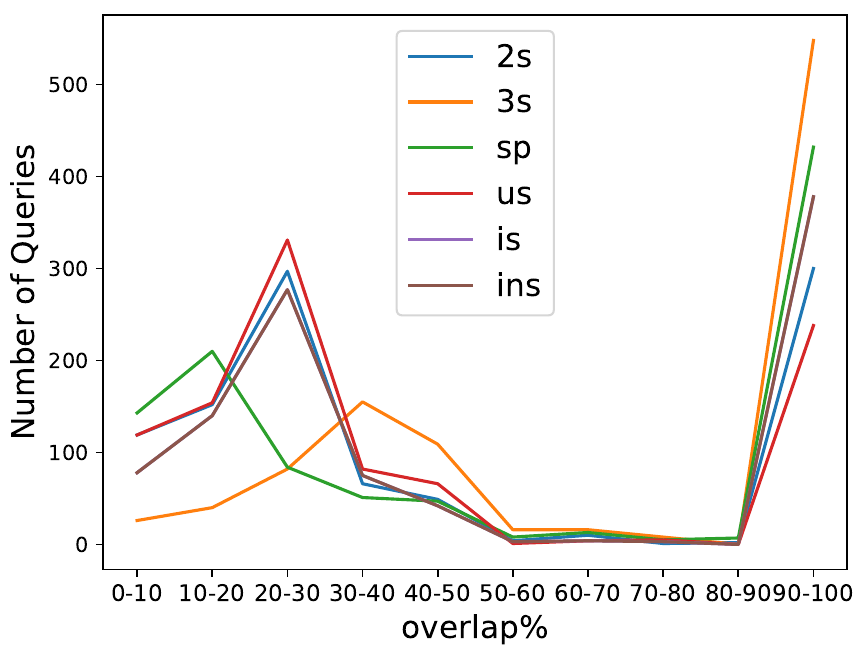}
    \caption{Overlap proportion between DAG and Tree-Form query answers from the NELL-DAG Easy test dataset.}
    \label{NELLQueryAns_Analysis}
\end{figure}

The model's ability to handle DAG queries is hard to evaluate when the answer sets of DAG and relaxed tree-form queries overlap significantly, as DAG query answers are a subset of those from relaxed tree-form queries. Specifically, a query embedding method can relax a DAG query into tree-form query by replacing all $\exists r\sqcap t. C$ with $\exists r. C \sqcap \exists t. C$.\footnote{We supplement the relaxed tree-form query types corresponding to proposed DAG types in Appendix \ref{DAGtoTF}.} This method performs well if the DAG query's answer set highly overlaps with that of the tree-form query. To test this, we analyzed random DAG queries from NELL \citep{NELL}. Figure \ref{NELLQueryAns_Analysis} shows that about $50\%$ of these DAG queries have answer sets with over $90\%$ overlap with tree-form queries. Similar analyses for FB15k and FB15k-237 are in Appendix \ref{QueryAns_Analysis}.

\begin{table*}[h]
    \centering
    \renewcommand{\arraystretch}{0.85} 
     \caption{The MRR performance of baseline models and our proposed version {\OurMethod} on easy benchmark datasets.}
      \vspace{-0.1cm}
    \begin{tabular}{llllllllll}
    \toprule
         \textbf{Dataset} & \textbf{Model} & \hspace{2mm}\textbf{2s} & \hspace{2mm}\textbf{3s} & \hspace{2mm}\textbf{sp} & \hspace{2mm}\textbf{is} & \hspace{2mm}\textbf{us} & \hspace{2mm}\textbf{Avg}$_{nn}$ & \hspace{2mm}\textbf{ins} & \hspace{2mm}\textbf{Avg}\\
    \midrule
          & Query2Box &  20.53&  34.03&  0.10&  23.31&  28.2 & \hspace{1mm} 21.23 & \hspace{3mm}-& \hspace{3mm}-\\

          & Query2Box ({\OurMethod}) & \textbf{37.61} $\uparrow$ & 49.42 $\uparrow$ & \textbf{41.71} $\uparrow$ & \textbf{40.57} $\uparrow$ & \textbf{42.75} $\uparrow$ & \hspace{2mm}\textbf{42.41} $\uparrow$ & \hspace{3mm}- & \hspace{3mm}-\\


          
          & BetaE & 15.30 & 28.78 & 13.10 & 17.72 & 16.69 & \hspace{1mm} 18.32 & 27.30 & 19.82\\

          NELL-DAG & BetaE ({\OurMethod}) & 36.87 $\uparrow$ & 57.14 $\uparrow$ & 34.95 $\uparrow$ & 39.90 $\uparrow$&  37.80 $\uparrow$& \hspace{1mm} 41.33 $\uparrow$ & \textbf{34.68} $\uparrow$ & \textbf{40.22} $\uparrow$\\


          
           & ConE & 23.55 & 39.38 & 19.48 & 25.28 & 25.01 & \hspace{1mm} 26.54 & 27.71 &  26.73 \\

          & ConE ({\OurMethod}) & 33.50 $\uparrow$ & \textbf{57.35} $\uparrow$ & 38.43 $\uparrow$ & 37.93 $\uparrow$ & 33.74 $\uparrow$ & \hspace{1mm} 40.19 $\uparrow$ & 33.94 $\uparrow$ & 39.15 $\uparrow$ \\


          

    \midrule
          & Query2Box & 6.84 & 11.61 & 9.26 & 6.48 & 4.61 & \hspace{2mm}7.76 & \hspace{3mm}- & \hspace{3mm}-\\

          & Query2Box ({\OurMethod}) & \textbf{7.41} $\uparrow$ & \textbf{12.64} $\uparrow$ & 10.07 $\uparrow$ & \textbf{7.32} $\uparrow$ & \textbf{5.03} $\uparrow$ & \hspace{2mm}8.49 $\uparrow$ & \hspace{3mm}- & \hspace{3mm}-\\


          
          & BetaE & 4.81 & 8.17  & 7.52 & 5.0 & 2.71 & \hspace{2mm}5.64 & 4.49 & 5.45 \\

          FB15k-237-DAG & BetaE ({\OurMethod}) & 6.27 $\uparrow$ & 12.11 $\uparrow$ & 9.64 $\uparrow$ & 6.66 $\uparrow$ &  4.09 $\uparrow$ & \hspace{2mm}7.75 $\uparrow$ & \textbf{6.58} $\uparrow$ & 7.56 $\uparrow$ \\


          
          & ConE & 4.90 & 9.21 & 8.88 & 5.52 & 3.08 & \hspace{2mm}6.32 & 4.80 & 6.06 \\

          & ConE ({\OurMethod}) & 6.87 $\uparrow$& 11.66 $\uparrow$& \textbf{12.36} $\uparrow$& 6.90 $\uparrow$ & 4.80 $\uparrow$& \hspace{2mm}\textbf{9.54} $\uparrow$& 6.08 $\uparrow$ & \textbf{8.12} $\uparrow$\\


          

    \midrule
          & Query2Box & 32.62 & 35.52 & 20.90 & 27.79 & 23.94 & \hspace{1mm} 28.15 & \hspace{3mm}- & \hspace{3mm}- \\

          & Query2Box ({\OurMethod}) & 37.74 $\uparrow$ & 42.93 $\uparrow$ & 24.30 $\uparrow$ & 29.37 $\uparrow$ & 25.97 $\uparrow$ & \hspace{1mm} 31.46 $\uparrow$ & \hspace{3mm}- & \hspace{3mm}- \\


          
          & BetaE & 25.91 & 33.13 & 28.20 & 22.21 & 23.31 & \hspace{1mm} 26.55 & 19.02 & 25.29\\

          FB15k-DAG & BetaE ({\OurMethod}) & 32.65 $\uparrow$  & 46.17 $\uparrow$  & 32.48 $\uparrow$ & 28.15 $\uparrow$  & 28.10 $\uparrow$ & \hspace{1mm} 33.50 $\uparrow$ & 25.39 $\uparrow$ & 32.15 $\uparrow$ \\


          
           & ConE & 32.10 & 37.42 & 32.37 & 27.14 & 27.85 & \hspace{1mm} 31.37 & 23.48 & 30.06\\

          & ConE ({\OurMethod}) & \textbf{41.67} $\uparrow$ & \textbf{56.70} $\uparrow$ & \textbf{33.36} $\uparrow$ & \textbf{36.54} $\uparrow$& \textbf{32.36} $\uparrow$ & \hspace{1mm} \textbf{40.12} $\uparrow$ & \textbf{30.86} $\uparrow$ & \textbf{38.58} $\uparrow$\\


          

    \bottomrule
    \end{tabular}
    \label{easy_main_results}
\end{table*}

\begin{table*}[h]
    \centering
\renewcommand{\arraystretch}{0.85} 
 \caption{The MRR performance of baseline models and our proposed version {\OurMethod} on hard benchmark datasets.}
  \vspace{-0.1cm}
    \label{hard_main_results}
    \begin{tabular}{llllllllll}
    \toprule
         \textbf{Dataset} & \textbf{Model} & \hspace{2mm}\textbf{2s} & \hspace{2mm}\textbf{3s} & \hspace{2mm}\textbf{sp} & \hspace{2mm}\textbf{is} & \hspace{2mm}\textbf{us} & \hspace{2mm}\textbf{Avg}$_{nn}$ & \hspace{2mm}\textbf{ins} & \hspace{2mm}\textbf{Avg}\\
    \midrule
          & Query2Box &  7.47 & 5.19 & 0.11 & 8.54 & 12.28 & \hspace{1mm} 6.72 & \hspace{3mm}- & \hspace{3mm}-\\
          
          & Query2Box ({\OurMethod}) &  25.38 $\uparrow$ & 20.13 $\uparrow$ & 21.25 $\uparrow$ & 24.85 $\uparrow$ & 29.24 $\uparrow$ & \hspace{1mm} 24.17 $\uparrow$ & \hspace{3mm}- & \hspace{3mm}-\\


          
          & BetaE & 14.38 & 16.23 & 7.99 & 13.32 & 13.03 & \hspace{1mm} 12.99 & 28.17 & 15.52\\

          NELL-DAG & BetaE ({\OurMethod}) & 27.68 $\uparrow$ & 32.25 $\uparrow$ & 16.36 $\uparrow$ & 26.14 $\uparrow$ & 29.19 $\uparrow$ & \hspace{1mm} 26.32 $\uparrow$ & 33.64 $\uparrow$ & 27.54 $\uparrow$\\


          
           & ConE &  20.31 & 18.88 & 12.02 & 19.59 & 22.31 & \hspace{1mm}18.62 & 29.45 & 20.43\\

          & ConE ({\OurMethod}) &  \textbf{30.71} $\uparrow$ & \textbf{38.41} $\uparrow$ & \textbf{24.76} $\uparrow$ & \textbf{28.44} $\uparrow$ & \textbf{31.06} $\uparrow$ & \hspace{1mm} \textbf{30.67} $\uparrow$ & \textbf{34.07} & \textbf{31.24} $\uparrow$ \\


          

    \midrule
          & Query2Box & 4.25 & 2.64 & 7.21 & 4.56 & 3.63 & \hspace{1mm}4.45 & \hspace{3mm}- & \hspace{3mm}- \\

          & Query2Box ({\OurMethod}) & 4.81 $\uparrow$ & \textbf{2.81} $\uparrow$ & 7.87 $\uparrow$ & \textbf{5.26} $\uparrow$ & \textbf{4.38} $\uparrow$ & \hspace{1mm}\textbf{6.95} $\uparrow$ & \hspace{3mm}- & \hspace{3mm}-\\


          
          & BetaE & 3.62 & 1.62 & 6.44 & 3.85 & 2.42 & \hspace{1mm}3.20 & 4.31 & 3.38\\

          FB15k-237-DAG & BetaE ({\OurMethod}) & \textbf{4.89} $\uparrow$ & 1.66 $\uparrow$ & 8.28 $\uparrow$ & 4.75 $\uparrow$ & 3.50 $\uparrow$ & \hspace{1mm}4.61 $\uparrow$ & \textbf{6.06} $\uparrow$ & 4.85$\uparrow$\\


          
           & ConE & 3.48 & 2.28 & 7.36 & 4.23 & 2.92 & \hspace{1mm}4.05 & 4.65 & 4.15 \\

          & ConE ({\OurMethod}) & 4.78 $\uparrow$ & 2.09 & \textbf{9.72} $\uparrow$ & 4.84 $\uparrow$  & 4.16 $\uparrow$  & \hspace{1mm}5.12 $\uparrow$ & 5.25 $\uparrow$ & \textbf{5.14} $\uparrow$\\


          

    \midrule
          & Query2Box & 31.86 & 33.32 & 18.46 & 25.59 & 22.59 & 26.36 & \hspace{3mm}- & \hspace{3mm}- \\

          & Query2Box ({\OurMethod}) & 33.78 $\uparrow$ & 39.67 $\uparrow$ & 19.61 $\uparrow$ & 26.91 $\uparrow$ & 24.76 $\uparrow$ & 28.95 $\uparrow$ & \hspace{3mm}- & \hspace{3mm}- \\



          & BetaE & 24.02 & 31.82 & 26.12 & 20.17 & 21.93 & 24.81 & 18.60 & 23.77 \\

          FB15k-DAG & BetaE ({\OurMethod}) & 30.57 $\uparrow$ & 44.30 $\uparrow$ & 29.35 $\uparrow$ & 25.72 $\uparrow$ & 26.63 $\uparrow$ & 31.31 $\uparrow$ & 25.18 $\uparrow$ & 30.29 $\uparrow$\\


          
           & ConE & 30.42 & 36.29 & \textbf{30.46} & 25.67 & 27.14 & 29.99 & 22.66 & 28.77\\

          & ConE ({\OurMethod}) & \textbf{40.14} $\uparrow$ & \textbf{57.06} $\uparrow$ & 29.23 & \textbf{34.63} $\uparrow$ & \textbf{31.45} $\uparrow$ & \textbf{38.50} $\uparrow$ & \textbf{30.74} $\uparrow$ & \textbf{37.21} $\uparrow$\\


          

    \bottomrule
    \end{tabular}
\end{table*}

To address this, we introduce two difficulty levels for DAG-QA benchmark datasets in Table \ref{Num:Test Queries}, \emph{\textbf{test-easy}} and \emph{\textbf{test-hard}}, such that
\begin{itemize}
    \item \emph{\textbf{Test-easy}} datasets are randomly generated, and the answer sets of some queries are probably highly similar to those of the corresponding tree queries.

    \item \emph{\textbf{Test-hard}} datasets are chosen from random queries where the overlap between the answer sets of these queries and their corresponding tree queries is below 0.5. For example, if the answer set of a DAG query is $\{a,b,c\}$ and that of its corresponding tree query is $\{a,b,c,d\}$ then this DAG query should be dismissed because the overlap ratio is 3/4.
\end{itemize}



\subsection{\textbf{RQ1:} How effective is {\OurMethod} for enhancing baseline models on DAG queries?}
\label{Main Results}
We assessed {\OurMethod}'s ability to extend tree-form query embedding methods to DAG queries by retraining and testing baseline models (Query2Box, BetaE, ConE) on new datasets, decomposing DAG queries into tree-form conjunctions. Implementation details are in Appendix \ref{computations of baseline models}. We applied {\OurMethod} to these models and reevaluated them on the same datasets under easy and hard test modes, using Mean Reciprocal Rank (MRR) as the metric. Given a query $\mathcal{Q}$, MRR represents the average of the reciprocal ranks of results, $\operatorname{MRR} = \frac{1}{|Q|}\sum_{i=1}^{|Q|}\frac{1}{\operatorname{rank}_{i}}$. 
\paragraph{\textbf{Main Results:}}
Tables \ref{easy_main_results} and \ref{hard_main_results} summarize the performance of the baseline methods, both with and without {\OurMethod}, under the easy and hard test modes. Based on these results, we draw the following conclusions. Firstly, the baseline methods show a significant performance drop from the easy to hard datasets due to the exclusion of "easy" DAG queries, as described in Section \ref{Query Generation}. This highlights the importance of developing datasets that effectively assess a model's ability to handle DAG queries, rather than just tree-form queries. Secondly, {\OurMethod} consistently delivers significant improvements to all baseline methods across all query types and datasets, in both easy and hard test modes. Specifically, {\OurMethod} significantly improves performance on the NELL-DAG dataset, with the average accuracy of baseline models nearly doubling when combined with {\OurMethod} compared to their standalone performance. Beyond these baseline models, we also implement other query embedding models, e.g., CQD\cite{CQD} and BiQE\cite{BiQE}, that are theoretically believed to be capable of handling DAG queries, on our new datasets. We perform comparison between the enhanced baselines and these methods in Appendix \ref{Comparison with other query embedding methods}. It demonstrates that {\OurMethod} can easily extend existing tree-form query embedding models to outperform these methods on DAG query datasets, further reinforcing its effectiveness.

\subsection{\textbf{RQ2:} How well does {\OurMethod} perform on the existing tree-form query benchmarks?}
An effective method for extending existing query embedding techniques to handle DAG queries should also ensure strong performance on the tree-form queries these methods were originally designed to process. Table \ref{DAGE-Tree-results(partial version)} presents the performance of the baseline models, as well as their performance when integrated with {\OurMethod}, on tree-form queries from NELL-QA \cite{BetaE}. The complete results on other two datasets are supplemented in Appendix \ref{DAGE-Tree-results(complete)}. {\OurMethod} enables these models to handle DAG queries while preserving their original performance on tree-form queries. Notably, {\OurMethod} significantly improves performance on DAG queries but has minimal impact on tree-form queries, confirming its effectiveness in enhancing baseline performance specifically for DAG query types.
\begin{table*}
    \centering
    \renewcommand{\arraystretch}{0.85} 
        \caption{The MRR performance of the retrained baseline models with {\OurMethod} method on tree-form query benchmark datasets}
        \vspace{-0.2cm}
    \label{DAGE-Tree-results(partial version)}
    \begin{tabular}{llllllllllll}
    \toprule
         \textbf{Dataset} & \textbf{Model} & \hspace{2mm}\textbf{1p} & \hspace{2mm}\textbf{2p} & \hspace{2mm}\textbf{3p} & \hspace{2mm}\textbf{2i} & \hspace{2mm}\textbf{3i} & \hspace{2mm}\textbf{pi} & \hspace{2mm}\textbf{ip} & \hspace{2mm}\textbf{2u} & \hspace{2mm}\textbf{up} & \textbf{Avg}\\
    \midrule
          & Query2Box &  42.7 & 14.5 & 11.7 & 34.7 & 45.8 & 23.2 & 17.4 & 12.0 & 10.7 & 23.6\\
          
          & Query2Box ({\OurMethod}) &  42.1  & 23.4  & 21.3  & 28.6 & 41.1 & 20.0 & 12.3 & 27.5 & 15.9 & 28.3 \\
          
          & BetaE & 53.0 & 13.0 & 11.4 & 37.6 & 47.5 & 24.1 & 14.3 & 12.2 & 8.5 & 24.6\\

          NELL-QA & BetaE ({\OurMethod}) & 53.4 & 12.9 & 10.8 & 37.6 & 47.1 & 23.8 & 13.8 & 12.3 & 8.3 & 24.4\\
          
           & ConE &  53.1 & 16.1 & 13.9 & 40.0 & 50.8 & 26.3 & 17.5 & 15.3 & 11.3 & 27.2 \\

          & ConE ({\OurMethod}) & 53.2 & 15.7 & 13.7 & 39.9 & 50.7 & 26.0 & 17.0 & 14.8 & 10.9 &  26.8 \\
          

    \bottomrule
    \end{tabular}
\end{table*}

\begin{table*}
    \centering
    \renewcommand{\arraystretch}{0.85} 
     \caption{The MRR performance of BetaE with {\OurMethod} on NELL-DAG hard benchmark dataset, along with its performance when integrated with additional logical constraints.}
       \vspace{-0.2cm}
    \label{Results: BetaE DAGE with logical constraints}
    \begin{tabular}{llllllllll}
    \toprule
          \textbf{Model} & \hspace{2mm}\textbf{2s} & \hspace{2mm}\textbf{3s} & \hspace{2mm}\textbf{sp} & \hspace{2mm}\textbf{is} & \hspace{2mm}\textbf{us} & \hspace{2mm}\textbf{Avg}$_{nn}$ & \hspace{2mm}\textbf{ins} & \hspace{2mm}\textbf{Avg}\\
          \midrule
           BetaE ({\OurMethod}) & 27.68 & 32.25 & 16.36 & 26.14 & 29.19 & \hspace{1mm} 26.32 & 33.64 & 27.54 \\

           BetaE ({\OurMethod}+Distr) & 27.91 &32.87&17.12&\textbf{27.02}&\textbf{30.13} &\hspace{1mm} 27.01&34.28&28.22\\

           BetaE ({\OurMethod}+Mono) & 28.01 & \textbf{33.56} & 16.89 & 26.93  & 29.47 &\hspace{1mm} 26.97 & 34.17 & 28.17\\

           BetaE ({\OurMethod}+Distr+Mono) & \textbf{28.11}	&  33.48&\textbf{17.67}&	26.83&29.36&\hspace{1mm} \textbf{27.09}&\textbf{34.49}&\textbf{28.32}\\
    \bottomrule
    \end{tabular}
\end{table*}

\subsection{\textbf{RQ3:} How do the logical constraints influence the performance of {\OurMethod}?}
Table \ref{Results: BetaE DAGE with logical constraints} summarizes the performances of BetaE \cite{BetaE} enhanced by DAGE with additional logical constraints, i.e., monoticity and restricted conjunction preservation in proposition \ref{prop:tautologies}. The performances of other baseline models, Query2Box and ConE, are provided in Table \ref{Complete Results: DAGE with logical constraints} of Appendix \ref{Performances of DAGE with additional logical constraints}. By examining these results, we are able to draw several key conclusions. First, it is evident that both monoticity and restricted conjunction preservation contribute to performance improvements across the board. Specifically, the enhancements brought about by monoticity regularization are more pronounced compared to those resulting from restricted conjunction preservation. Furthermore, we find that the combination of both logical constraints consistently enhances DAGE's performance on DAG query answering tasks, highlighting the importance of incorporating constraints in complex query answering.

\begin{figure}
    \centering
    \includegraphics[width=0.73\linewidth]{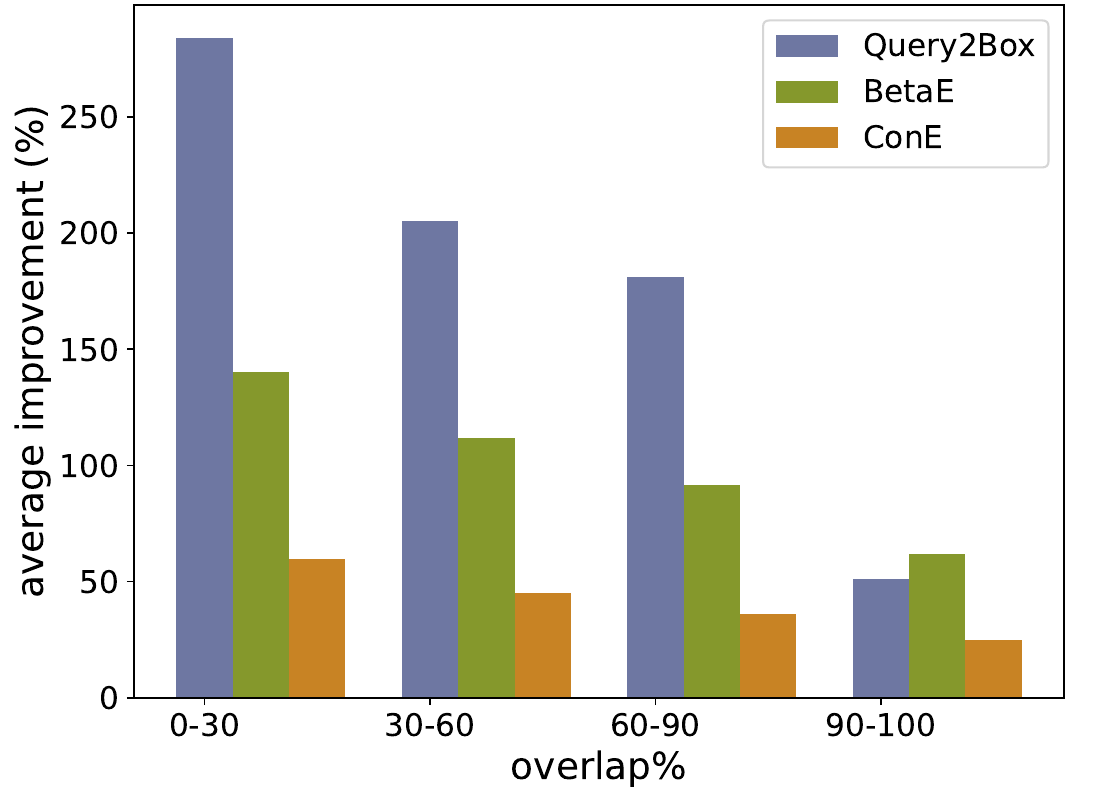}
    \caption{Average percentage of improvement from baseline model to baseline (DAGE) for different overlap ratios}
    \label{ablation study}
    \vspace{-10pt}
\end{figure}

\subsection{Ablation study}
\label{section:ablation study}
To more effectively examine the specific impact of DAGE on baseline models to DAG queries, a detailed analysis was conducted on them based on the NELL-DAG query answering dataset. We divide the easy test dataset into four groups, $0-30\%$, $30-60\%$, $60-90\%$, and $90-100\%$, based on the overlap ratio between DAG query answers and their corresponding tree-form query answers. The specific number of queries in each category can be found in Figure \ref{NELLQueryAns_Analysis}. Figure \ref{ablation study} shows the average performance improvement, in percentage, of the baseline models when enhanced with DAGE across the subgroups of test queries. Most of DAGE's improvements occur in queries with lower answer overlap ratios. For higher overlap ratios, DAGE bring less improvement. This shows that DAGE significantly improves baseline models, particularly on challenging tasks they previously couldn't handle on their own.

\section{Related Work}
\emph{Query Embedding Methods.} Path-based \cite{deeppath,LinSX18}, neural \cite{GQE,querytobox,ConE,BetaE,BiQE}, and neural-symbolic \cite{CQD,GNNQE,FuzzQE} methods have been developed to answer (subsets of) queries. Among these methods, geometric and probabilistic query embedding approaches \cite{GQE,querytobox,ConE,BetaE,PConE} provide an effective way to answer tree-form queries over incomplete and noisy KGs. These methods achieve this by representing sets of entities as geometric shapes or probability distributions, such as boxes \cite{querytobox}, cones \cite{ConE}, or Beta distributions \cite{BetaE}, and applying neural logic operations directly on these representations. The Graph Query Embedding (GQE) method \cite{GQE} was one of the earliest approaches, designed to handle only conjunctive queries by representing the query $q$ as a single vector  using neural translational operations. However, representing a query as a single vector limits its ability to effectively capture multiple entities. Query2Box \cite{querytobox} addresses this limitation by representing entities as points within boxes, enabling it to model the intersection of entity sets as the intersection of boxes in vector space. ConE \cite{ConE} introduced the first geometry-based query embedding approach capable of handling negation by embedding entity sets (or query embeddings) as cones in Euclidean space. However, these established theories and methods are limited to tree-form queries. There is a lack of techniques that can extend their application to DAG queries.

\emph{Ontology Embedding.} Another line of research focuses on applying geometric embeddings to encode concepts and roles in ontologies \cite{chen2024ontology, XiongGeometric}. This approach shares a fundamental similarity with query embedding methods, as both use geometric representations—such as boxes \cite{xiong2022faithful, zhu2023towards, zhu2024approx} or n-balls \cite{DBLP:journals/corr/abs-2202-14018}—to model sets of entities. Additionally, they leverage transformation functions, such as affine transformations \cite{xiong2022faithful}, or dual-box \cite{box2el} representations, to capture the semantics of roles and relationships effectively.


\section{Conclusion}

In this paper, we introduce DAG queries, a broader class of queries. We propose \OurMethod, a plug-and-play module that extends existing tree-form query embedding approaches to handle DAG queries, whose computation graphs contain more than one paths between two nodes. {\OurMethod}
handle this issue by merging the possible multiple paths through a relational combinator, which corresponds to the conjunction operator of relations in $\ALCOIR$). 
We propose proper regularization terms to encourage the inference of query embeddings to satisfy desired tautologies including monotonicity and restricted conjunction preserving. 
We create novel benchmarks consisting of DAG queries for evaluating DAG query embedding approaches. 
We implementDAGE upon three existing query embedding approaches, and the results show that {\OurMethod} significantly outperforms its corresponding counterpart on DAG queries while maintaining competitive performance on tree-form queries.  

One limitation of {\OurMethod} is its inability to enforce hard constraints on tautologies, as the regularization loss cannot reach zero. Future work will explore embedding approaches that inherently respect these tautologies without regularization. Another limitation is that many test queries can be simplified, requiring inference on only a subset of links, as shown in \citep{gregucci2024complex}. We plan to test our approach on "truly complex" queries requiring inference on all links \citep{gregucci2024complex}.

\begin{acks} 
The authors thank the International Max Planck Research School for Intelligent Systems for supporting Yunjie He and Yuqicheng Zhu. 
The work was partially supported by EU Projects Graph Massiviser (GA 101093202), Dome 4.0 (GA 953163), enRichMyData (GA 101070284) and SMARTY (GA 101140087), and the German Research Foundation, DFG (GA SFB-1574-471687386).
\end{acks}

\clearpage
\bibliographystyle{unsrt}
\bibliography{reference}
\balance
\appendix
\section{Specific details of computations of baseline models with DAGE}
\label{computations of baseline models}
In this section, we supplement the specific details of the computations, i.e., relational transformation, intersection operator and complement operator, of the query embedding models involved in our experiments. Note that we do not introduce the union operator because queries involving with union can be translated into disjunctive normal form (DNF), more details can be checked in \cite{querytobox}.

\subsection{Query2Box}
Query2Box models concepts in the vector space using boxes (i.e., axis-aligned hyper-rectangles) and defines a box in $\mathbb{R}^d$ by $\mathbf{p}=(\operatorname{Cen}(\mathbf{p}), \operatorname{Off}(\mathbf{p})) \in \mathbb{R}^{2 d}$ as

\begin{equation}
\operatorname{Box}_{\mathbf{p}} \equiv\left\{\mathbf{v} \in \mathbb{R}^d: \operatorname{Cen}(\mathbf{p})-\operatorname{Off}(\mathbf{p}) \preceq \mathbf{v} \preceq \operatorname{Cen}(\mathbf{p})+\operatorname{Off}(\mathbf{p})\right\}
\end{equation}

where $\preceq$ is element-wise inequality, $\operatorname{Cen}(\mathbf{p}) \in \mathbb{R}^d$ is the center of the box, and $\operatorname{Off}(\mathbf{p}) \in \mathbb{R}_{\geq 0}^d$ is the positive offset of the box, modeling the size of the box. 

The operations of concepts can be defined by

\begin{itemize}
    \item \textbf{Relational Transformation} maps from one box to another box using a box-to-box translation. This is achieved by translating the center and getting a larger offset. This is modeled by $\mathbf{p}+\mathbf{r}$, where each relation r is associated with a relation embedding $\mathbf{r} = (\text{Cen}(\mathbf{r}), \text{Off}(\mathbf{r})) \in \mathbb{R}^{2 d}$.

    \item \textbf{Intersection Operator} models the intersection of a set of box embedding $\{\mathbf{p}_{1}, \cdots, \mathbf{p}_{n}\}$ as $\mathbf{p}_{inter}$ = $(\text{Cen}$ $(\mathbf{p}_{inter}),$ $\text{Off}(\mathbf{p}_{inter}))$, such that
    \begin{equation}
        \Ce{\Em{p_{{\rm inter}}}} = \sum_i \mathbf{w}_i \odot \Ce{\Em{p_i}}, 
    \end{equation}
    where \[ \mathbf{w}_i = \frac{\exp(\textsc{MLP}(\Em{p_i}))}{\sum_j {\exp (\textsc{MLP}(\Em{p_j}))}},\]
    \begin{equation}
    \begin{split}
        \Of{\Em{p_{{\rm inter}}}} = {\rm Min}(\{\Of{\Em{p_1}}, \dots, \Of{\Em{p_n}}\}) \\
        \odot \sigma({\rm DeepSets}(\{\Em{p_1}, \dots, \Em{p_n}\})),
    \end{split}
    \end{equation}
    where $\odot$ is the dimension-wise product, ${\rm MLP}(\cdot): \mathbb{R}^{2d} \to \mathbb{R}^{d}$ is the Multi-Layer Perceptron, $\sigma(\cdot)$ is the sigmoid function, ${\rm DeepSets}(\cdot)$ is the permutation-invariant deep architecture \citep{deepsets}, and both ${\rm Min}(\cdot)$ and ${\exp}(\cdot)$ are applied in a dimension-wise manner.
    
    \item \textbf{Distance function}
    Given a query box $\Em{p} \in \mathbb{R}^{2d}$ and an entity vector $\Em{a} \in \mathbb{R}^{d}$, their distance is define as
    \begin{align} 
    {\rm dist}_{\rm box}(\Em{a};\Em{p}) = {\rm dist}_{\rm outside}(\Em{a};\Em{p}) + \alpha \cdot {\rm dist}_{\rm inside}(\Em{a};\Em{p}),
    \end{align}
    where $\Em{q_{\rm max}} = \Ce{\Em{p}} + \Of{\Em{p}} \in \mathbb{R}^d$, $\Em{p_{\rm min}} = \Ce{\Em{p}} - \Of{\Em{p}} \in \mathbb{R}^d$ and $0 < \alpha < 1$ is a fixed scalar, and 
        \begin{equation}
        {\rm dist}_{\rm outside}(\Em{a};\Em{p}) = \| {\rm Max}(\Em{a} - \Em{q_{\rm max}}, \Em{0}) + \\ {\rm Max}(\Em{q_{\rm min}} - \Em{a}, \Em{0}) \|_1,    
        \end{equation}
        \begin{equation}
        {\rm dist}_{\rm inside}(\Em{a};\Em{p}) = \|\Ce{\Em{p}} - {\rm Min}(\Em{q_{\rm max}}, {\rm Max}(\Em{q_{\rm min}}, \Em{a}) ) \|_1.
        \end{equation}
    
    \item \textbf{Insideness function}
    Given query box embeddings $\mathbf{p}_1$ and $\mathbf{p}_2$, the Query2Box insideness function measures if $\mathbf{p}_1$ is inside $\mathbf{p}_2$ by returning the overlap ratio between their intersection and $\mathbf{p}_1$. A higher ratio indicates a greater likelihood that $\mathbf{p}_1$ is inside $\mathbf{p}_2$. In this case, the insideness function is defined as below:
    \begin{equation}
        \text{Insideness}(\mathbf{p}_{1}, \mathbf{p}_{2}) =  \frac{\text{BoxVolume}(\text{Intersect}(\mathbf{p}_{1}, \mathbf{p}_{2}))}{\text{BoxVolume}(\mathbf{p}_{1})}
    \end{equation}
    where BoxVolume($\mathbf{p}$) measures the volume of the box embedding via a softplus function, such that 
    \[\text{BoxVolume} = \underset{i}{\Pi}\frac{1}{\beta} log(1+\text{exp}(\beta \cdot \text{Off}(\mathbf{p})_{i})),\] 
    \[\text{Off}(\mathbf{p})_{i} \in \text{Off}(\mathbf{p})\]

    \item \textbf{Difference function}
    Given two boxes, $\mathbf{p}_1$ and $\mathbf{p}_2$, their difference is modeled as 
    \begin{multline}
        \text{Diff}(\mathbf{p}_{1}, \mathbf{p}_{2}) = \underset{i}{\sum}\mid \text{Cen}(\mathbf{p}_{1,i}) -\text{Cen}(\mathbf{p}_{2,i})\mid + \\ \mid \text{Off}(\mathbf{p}_{1,i}) -\text{Off}(\mathbf{p}_{2,i})\mid
    \end{multline}
\end{itemize}



\subsection{BetaE}
BetaE represents concepts by the Cartesian product of multiple Beta distributions: $\embedding{C}=\left[\left(\alpha_1, \beta_1\right), \ldots,\left(\alpha_n, \beta_n\right)\right]$ where each component is a Beta distribution $\operatorname{Beta}\left(\alpha, \beta\right)$  controlled with two shape parameters $\alpha$ and $\beta$. 

\begin{itemize}
    \item \textbf{Relational Transformation} maps from one Beta embedding $\mathbf{S}$ to another Beta embedding $\mathbf{S}^{\prime}$ given the relation type $r$. This is modeled by a transformation neural network for each relation type $r$ using a multi-layer perceptron (MLP):

\begin{equation}
    \RelT_r(\embedding{C}) = \operatorname{MLP}_r(\embedding{C})
\end{equation}

    \item \textbf{Intersection Operator} is modeled by taking the weighted product of the PDFs of the input Beta embeddings 

\begin{equation}
 \Intersection(\embedding{C_1}, \cdots, \embedding{C_n}) = 
 \frac{1}{Z} \prod p_{\embedding{C}}^{w_1} \cdots p_{\embedding{C}}^{w_n} 
\end{equation}
where $Z$ is a normalization constant and $w_1, \cdots, w_n$ are the weights with their sum equal to $1$.

    \item \textbf{Complement Operator} is modeled by taking the reciprocal of the shape parameters.

    \begin{equation}
        \Complement(\embedding{C}) = \left[\left(\frac{1}{\alpha_1} , \frac{1}{\beta_1}\right), \ldots,\left(\frac{1}{\alpha_n}, \frac{1}{\beta_n}\right)\right]
    \end{equation}

    \item \textbf{Distance function} Given an answer entity embedding $\mathbf{a}$ with parameters $\left[\left(\alpha_1^a, \beta_1^a\right), \ldots,\left(\alpha_n^a, \beta_n^a\right)\right]$, and a query embedding $\mathbf{q}$ with parameters $\left[\left(\alpha_1^q, \beta_1^q\right), \ldots,\left(\alpha_n^q, \beta_n^q\right)\right]$, we define the distance between this entity $a$ and the query $q$ as the sum of KL divergence between the two Beta embeddings along each dimension:
\begin{equation}
    \operatorname{dist_{beta}}(a; q)=\sum_{i=1}^n \mathrm{KL}\left(p_{\mathbf{a}, \mathbf{i}} ; p_{\mathbf{q}, \mathbf{i}}\right)
\end{equation}
    
    \item \textbf{Difference function}
    Given two beta query embeddings, \(\embedding{C_1}=\left[\left(\alpha_{1\textbf{1}}, \beta_{1\textbf{1}}\right), \ldots,\left(\alpha_{1\textbf{n}}, \beta_{1\textbf{n}}\right)\right]\) and  \\
    \(\embedding{C_2}=\left[\left(\alpha_{\textbf{21}}, \beta_{\textbf{21}}\right), \ldots,\left(\alpha_{\textbf{2n}}, \beta_{\textbf{2n}}\right)\right]\), their difference is modeled as 
    \begin{multline}
        \text{Diff}_{beta}(\embedding{C_1}, \embedding{C_2}) = \underset{i\in \{1,\cdots,n\}}{\sum}\mid \alpha_{\textbf{1i}} - \alpha_{\textbf{2i}}\mid + \\ \mid \beta_{\textbf{1i}} - \beta_{\textbf{2i}} \mid
    \end{multline}

    \item \textbf{insideness function} Given query beta embeddings $\mathbf{q}_1$ and $\mathbf{q}_2$, the BetaE insideness function meansures if $\mathbf{q}_1$ is inside $\mathbf{q}_2$ by returning the difference between their intersection and $\mathbf{q}_1$. Their intersection is expected to match $\mathbf{q}_1$ if  $\mathbf{q}_1$ is fully inside $\mathbf{q}_2$. Thus, the beta insideness function can be defined as 
    \begin{equation}
        \text{Insideness}_{beta} = \text{Diff}_{beta}(\text{Intersect}_{beta}(\mathbf{q}_1, \mathbf{q}_2), \mathbf{q}_1)
    \end{equation}

\end{itemize}

\subsection{ConE}

ConE model concepts by a Cartesian product of sector-cones. Specifically, ConE uses the parameter $\theta_{\mathrm{ax}}^i$ to represent the semantic center, and the parameter $\theta_{\mathrm{ap}}^i$ to determine the boundary of the query. Given a $d$-ary Cartesian product, the embedding of concept is defined as

\begin{equation}
\embedding{C} = \operatorname{MultiCone}\left(\boldsymbol{\theta}_{\mathrm{ax}}, \boldsymbol{\theta}_{\mathrm{ap}}\right)
\end{equation}
where $\boldsymbol{\theta}_{\mathrm{ax}} \in[-\pi, \pi)^d$ are axes and $\boldsymbol{\theta}_{\mathrm{ap}} \in[0,2 \pi]^d$ are apertures.

\begin{itemize}

    \item \textbf{Nominal} is defined as a cone with apertures $0$.
    \begin{equation}
        \Nominal(\embedding{a}) = \operatorname{MultiCone}(\boldsymbol{\theta}_{\mathrm{ax}},0)
    \end{equation}
    \item \textbf{Relational Transformation} maps a cone embedding to another cone embedding. This is implemented by a relation specific transformation. 


 \begin{equation}
\RelT(_r\left(\operatorname{MultiCone}(\boldsymbol{\theta}_{\mathrm{ax}},\boldsymbol{\theta}_{\mathrm{ap}})\right)=g\left(\mathbf{M L P}\left(\left[\boldsymbol{\theta}_{\mathrm{ax}}+\boldsymbol{\theta}_{\mathrm{ax}, r} ; \boldsymbol{\theta}_{\mathrm{ap}}+\boldsymbol{\theta}_{\mathrm{ap}, r}\right]\right)\right)
\end{equation}

    \item \textbf{Intersection Operator} Suppose that $\embedding{C}=\left(\boldsymbol{\theta}_{\mathrm{ax}}, \boldsymbol{\theta}_{\mathrm{ap}}\right)$ and $\embedding{C_i}=\left(\boldsymbol{\theta}_{i, \mathrm{ax}}, \boldsymbol{\theta}_{i, \mathrm{ap}}\right)$ are cone embeddings for $C$ and $C_i$, respectively. We define the intersection operator as follows:
\begin{equation}
\begin{aligned}
& \boldsymbol{\theta}_{\mathrm{ax}}=\operatorname{SemanticAverage}\left(\mathbf{V}_{q_1}^c, \ldots, \mathbf{V}_{q_n}^c\right) \\
& \boldsymbol{\theta}_{\mathrm{ap}}=\operatorname{CardMin}\left(\mathbf{V}_{q_1}^c, \ldots, \mathbf{V}_{q_n}^c\right)
\end{aligned}
\end{equation}

where SemanticAverage $(\cdot)$ and CardMin( $\cdot$ ) generates semantic centers and apertures, respectively.
    
    \item \textbf{Complement Operator} Suppose that $\embedding{C} = \operatorname{MultiCone}$ $\left(\boldsymbol{\theta}_{\mathrm{ax}}, \boldsymbol{\theta}_{\mathrm{ap}}\right)$ and $\embedding{\neg C}=\operatorname{MultiCone}\left(\boldsymbol{\theta}_{\mathrm{ax}}^{\prime}, \boldsymbol{\theta}_{\mathrm{ap}}^{\prime}\right)$. We define the complement operator as:

\begin{equation}
\begin{aligned}
& {\left[\boldsymbol{\theta}_{\mathrm{ax}}^{\prime}\right]_i= \begin{cases}{\left[\boldsymbol{\theta}_{\mathrm{ax}}\right]_i-\pi,} & \text { if }\left[\boldsymbol{\theta}_{\mathrm{ax}}\right]_i \geq 0 \\
{\left[\boldsymbol{\theta}_{\mathrm{ax}}\right]_i+\pi,} & \text { if }\left[\boldsymbol{\theta}_{\mathrm{ax}}\right]_i<0\end{cases} } \\
& {\left[\boldsymbol{\theta}_{\mathrm{ap}}^{\prime}\right]_i=2 \pi-\left[\boldsymbol{\theta}_{\mathrm{ap}}\right]_i .}
\end{aligned}
\end{equation}

    \item \textbf{Distance function.} 
    Suppose that the entity embedding is $\textbf{v}=(\gb{\theta}^v_{\taxis},\textbf{0})$, and the query cone embedding is $\textbf{V}_q^c=(\gb{\theta}_{\taxis},\gb{\theta}_{\targ})$, $\gb{\theta}_L=\gb{\theta}_{\taxis}-\gb{\theta}_{\targ}/2$ and $\gb{\theta}_U=\gb{\theta}_{\taxis}+\gb{\theta}_{\targ}/2$. The distance between the query and the entity is defined as
    \begin{align}
    d_{con}(\ee;\qe^c)=d_o(\ee;\qe^c)+\lambda d_i(\ee;\qe^c).
    \end{align}
    The outside distance and the inside distance are 
    \begin{align*}
    d_o&=\left\|\min\left\{\left|\sin\left(\gb{\theta}^v_{\taxis}-\gb{\theta}_L\right)/2\right|,\left|\sin\left(\gb{\theta}^v_{\taxis}-\gb{\theta}_U\right)/2\right|\right\}\right\|_1,\\
    d_i&=\left\|\min\left\{\left|\sin\left(\gb{\theta}^v_{\taxis}-\gb{\theta}_{\taxis}\right)/2\right|,\left|\sin\left(\gb{\theta}_{\targ}\right)/2\right| \right\}\right\|_1,
    \end{align*}
where $\|\cdot\|_1$is the $L_1$ norm, $\sin(\cdot)$ and $\min(\cdot)$ are element-wise sine and minimization functions.
    
    \item \textbf{Difference function}
    Given two cone embeddings, $\embedding{C1} = \operatorname{MultiCone}\left(\boldsymbol{\theta}_{\mathrm{ax,1}}, \boldsymbol{\theta}_{\mathrm{ap,1}}\right)$ and $\embedding{C2} = \operatorname{MultiCone}\left(\boldsymbol{\theta}_{\mathrm{ax,2}}, \boldsymbol{\theta}_{\mathrm{ap,2}}\right)$, their difference is modeled as 
    \begin{multline}
        \text{Diff}_{cone}(\embedding{C_1}, \embedding{C_2}) = \underset{i\in \{1,\cdots,n\}}{\sum}\mid \theta_{\mathrm{ax},\textbf{1i}} - \theta_{\mathrm{ax},\textbf{2i}}\mid + \\ \mid \theta_{\mathrm{arg},\textbf{1i}} - \theta_{\mathrm{arg},\textbf{2i}} \mid
    \end{multline}

    \item \textbf{insideness function}
    Given query cone embeddings $\mathbf{q}_1$ and $\mathbf{q}_2$, the ConE insideness function meansures if $\mathbf{q}_1$ is inside $\mathbf{q}_2$ by returning the difference between their intersection and $\mathbf{q}_1$. Their intersection is expected to match $\mathbf{q}_1$ if  $\mathbf{q}_1$ is fully inside $\mathbf{q}_2$. Thus, the ConE insideness function can be defined as 
    \begin{equation}
        \text{Insideness}_{cone} = \text{Diff}_{cone}(\text{Intersect}(\mathbf{q}_1, \mathbf{q}_2), \mathbf{q}_1)
    \end{equation}

\end{itemize}

\section{Interpretation of $\ALCOIR$ Descriptions}%
\label{sec:ACOIR-descriptions}

Table~\ref{table:interpretation-descriptions} presents how $\ALCOIR$ concept and role descriptions are interpreted. Given an interpretation $\Int$, for each description $X$ the interpretation of the description $X^\Int$ is recursively defined.

\begin{table}[h]
  \caption{Interpretation of $\ALCOIR$ Descriptions. On the left are concept or role descriptions $X$, and on the right are the interpretations $X^\Int$.}%
  \label{table:interpretation-descriptions}
  \centering
  \begin{tabular}{ll}
    \toprule
    $X$
    & $X^\Int$ \\ \midrule
    $\top$
    & $\Delta^\Int$ \\
    $\nominal{a}$
    & $\{ a^\Int \}$ \\
    $C \sqcap D$
    & $C^{\Int} \cap D^{\Int}$ \\
    $\neg C$
    & $\Delta^\Int \setminus C^{\Int}$ \\
    $\exists R.C$
    & $\{ u \mid (u,v) \in R^\Int \text{ and } v \in C^\Int\}$ \\
    \midrule
    $R^-$
    & $\{(u,v) \mid (v,u) \in R^\Int \}$ \\
    $R \circ S$
    & $\{(u,v) \mid \text{exists } w, (u,w) \in R^\Int \text{ and } (w,v) \in S^\Int \}$ \\
    $R \sqcap S$
    & $\{(u,v) \mid (u,v) \in R^\Int \text{ and } (u,v) \in S^\Int \}$ \\
    $R^+$
    & $\bigcup_{i \geq 1} (R^\Int)^i$, i.e., $R^+$ is the transitive closure of $R^I$ \\
    \bottomrule
  \end{tabular}
\end{table}

\section{Computation Graphs of DAG queries}%
\label{sec:computation-graphs-appendix}

In this appendix, we define a graph representation for DAG queries.

\begin{definition}[Computation Graph Role Composition]
  The \emph{role composition} of a computation graph $\ComputationGraph = (\ComputationNodes, \ComputationEdges,\ComputationNodeLabel, \ComputationTarget)$ with a role description $R$, denoted $\RoleComposition{\ComputationGraph}{E}$, is the computation graph $(\ComputationNodes', \ComputationEdges',\ComputationNodeLabel', \ComputationTarget')$ defined recursively as follows:
  \begin{enumerate}
      \item If the role description $R$ is a role name $r \in \RoleNames$ or the inverse $r^-$ of a role name $r$ then:
      \begin{enumerate}
          \item $\ComputationNodes' = \ComputationNodes \cup \{u\}$ where $u \notin \ComputationNodes$;
          \item $\ComputationEdges' = \ComputationEdges \cup \{(\ComputationTarget, u)\}$;
          \item $\ComputationNodeLabel' = \ComputationNodeLabel \cup \{ u \mapsto \exists R\}$; and
          \item $\ComputationTarget' = u$.
      \end{enumerate}
    \item $\RoleComposition{\ComputationGraph}{R \circ S} = \RoleComposition{\RoleComposition{\ComputationGraph}{R}}{S}$.
    \item $\RoleComposition{\ComputationGraph}{R^{--}} = \RoleComposition{\ComputationGraph}{R}$.
    \item $\RoleComposition{\ComputationGraph}{(R \circ S)^-} = \RoleComposition{\ComputationGraph}{S^- \circ R^-}$.
    \item $\RoleComposition{\ComputationGraph}{(R \sqcap S)^-} = \RoleComposition{\ComputationGraph}{R^- \sqcap S^-}$.
    \item Let $R_1$ and $R_2$ be two role descriptions, $\RoleComposition{\ComputationGraph}{R_1}$ be $(\ComputationNodes_1, \ComputationEdges_1, \ComputationNodeLabel_1, \ComputationTarget_1)$ and $\RoleComposition{\ComputationGraph}{R_2}$ be $(\ComputationNodes_2, \ComputationEdges_2, \ComputationNodeLabel_2, \ComputationTarget_2)$. If $R$ is $R_1 \sqcap R_2$ then:
    \begin{enumerate}
        \item $\ComputationNodes' = \ComputationNodes_1 \cup \ComputationNodes_2 \cup \{u\}$ where $\ComputationNodes_1 \cap \ComputationNodes_2 = \ComputationNodes$ and $u \notin \ComputationNodes_1 \cup \ComputationNodes_2$;
        \item $\ComputationEdges' = \ComputationEdges_1 \cup \ComputationEdges_2 \cup \{(\ComputationTarget_1, u), (\ComputationTarget_2, u)\}$;
        \item $\ComputationNodeLabel' = \ComputationNodeLabel_1 \cup \ComputationNodeLabel_2 \cup \{ u \mapsto \sqcap \}$;
        \item $\ComputationTarget' = u$.
    \end{enumerate}
  \end{enumerate}
\end{definition}

\begin{definition}[DAG computation graph]     
  The \emph{computation graph} of a DAG query $\Query$ is the smallest computation graph $\ComputationGraph(C) = (\ComputationNodes(\Query), \ComputationEdges(\Query),\ComputationNodeLabel(\Query), \ComputationTarget(\Query))$ defined recursively as follows:
  \begin{enumerate}
    \item If $\Query$ is $\nominal{a}$ then:
    \begin{enumerate}
        \item $\ComputationNodes(\Query) = \{u\}$;
        \item $\ComputationEdges(\Query) = \emptyset$;
        \item $\ComputationNodeLabel(\Query) = \{u \mapsto \nominal{a}\}$; and
        \item $\ComputationTarget(\Query) = u$.
    \end{enumerate}
    \item If $\Query$ is $C \sqcap D$ then:
    \begin{enumerate}
        \item $\ComputationNodes(\Query) = \ComputationNodes(C) \cup \ComputationEdges(D) \cup \{u\}$ where the sets $\ComputationNodes(C)$, $\ComputationEdges(D)$, and $\{u\}$ are pairwise disjoint.
        \item $\ComputationEdges(\Query) = \ComputationEdges(C) \cup \ComputationEdges(D) \cup \{(\ComputationTarget(C), u), (\ComputationTarget(D), u)\}$
        \item $\ComputationNodeLabel(\Query) = \ComputationNodeLabel(C) \cup \ComputationNodeLabel(D) \cup \{u \mapsto \sqcap\}$.
        \item $\ComputationTarget(Q) = u$.
    \end{enumerate}
    \item If $\Query$ is $\neg C$ then:
    \begin{enumerate}
        \item $\ComputationNodes(\Query) = \ComputationNodes(C) \cup \{u\}$ where $\ComputationNodes(C) \cap \{u\} = \emptyset$.
        \item $\ComputationEdges(Q) = \ComputationEdges(C) \cup \{(\ComputationTarget(C), u)\}$.
        \item $\ComputationNodeLabel(Q) = \ComputationNodeLabel(C) \cup \{ u \mapsto \neg\}$.
        \item $\ComputationTarget(Q) = u$.
    \end{enumerate}
    \item If $\Query$ is $\exists R.C$ then $\ComputationGraph(Q) = \RoleComposition{\ComputationGraph(C)}{R}$.
  \end{enumerate}
\end{definition}

\section{Relaxed tree-form query types}
\begin{figure}
    \centering
    \includegraphics[width=\linewidth]{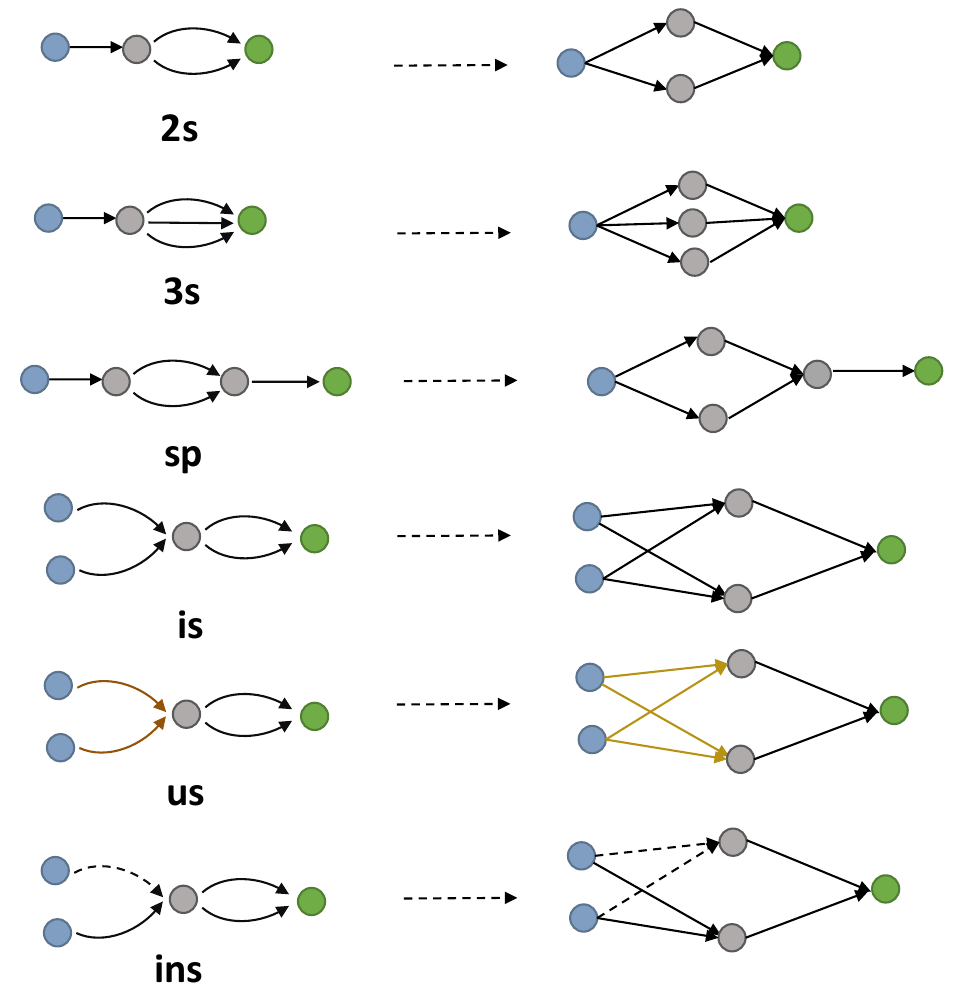}
    \caption{Transformation from DAG query to relaxed tree-form query}
    \label{Fig:DAGtoTF}
\end{figure}

Figure \ref{Fig:DAGtoTF} illustrates the query graphs of the new DAG query types and their corresponding relaxed tree-form query types.
\label{DAGtoTF}

Each of these queries are expressed as $\ALCOIR$ concepts as follows:
\begin{align}
  2s &\Coloneqq \exists (r_1 \circ (r_2 \sqcap r_3))^-.\nominal{e_1}, \\
  \Approximated(2s) &\Coloneqq 
  \begin{aligned}[t]
    & \exists (r_1 \circ r_2)^-.\nominal{e_1}\; \sqcap\\
    & \exists (r_1 \circ r_3)^-.\nominal{e_1},
  \end{aligned} \\
  3s &\Coloneqq \exists (r_1 \circ (r_2 \sqcap r_3 \sqcap r_4))^-.\nominal{e_1}, \\
  \Approximated(3s) &\Coloneqq
  \begin{aligned}[t]
  &\exists (r_1 \circ r_2)^-.\nominal{e_1}\; \sqcap\\
  &\exists (r_1 \circ r_3)^-.\nominal{e_1}\; \sqcap\\
  &\exists (r_1 \circ r_4)^-.\nominal{e_1},
  \end{aligned}\\
  sp &\Coloneqq \exists (r_1 \circ (r_2 \sqcap r_3) \circ r_4)^-.\nominal{e_1}, \\
  \Approximated(sp) &\Coloneqq 
  \begin{aligned}[t]
    & \exists (r_1 \circ r_2 \circ r_4)^-.\nominal{e_1}\; \sqcap\\
    & \exists (r_1 \circ r_3 \circ r_4)^-.\nominal{e_1},
  \end{aligned} \\
  is &\Coloneqq \exists (r_3 \sqcap r_4)^-.(\exists r_1\nominal{e_1} \sqcap \exists r_2\nominal{e_2}), \\
  \Approximated(is) &\Coloneqq 
  \begin{aligned}[t]
    & \exists r_3^-.(\exists r_1.\nominal{e_1} \sqcap \exists r_2 \nominal{e_2})\; \sqcap\\
    & \exists r_4^-.(\exists r_1 \nominal{e_1} \sqcap \exists r_2 \nominal{e_2}),
  \end{aligned}
\end{align}
\begin{align}
  us &\Coloneqq \exists (r_3 \sqcap r_4)^-.(\exists r_1\nominal{e_1} \sqcup \exists r_2\nominal{e_2}), \\
  \Approximated(us) &\Coloneqq 
  \begin{aligned}[t]
    & \exists r_3^-.(\exists r_1.\nominal{e_1} \sqcup \exists r_2 \nominal{e_2})\; \sqcap\\
    & \exists r_4^-.(\exists r_1 \nominal{e_1} \sqcup \exists r_2 \nominal{e_2}),
  \end{aligned}\\
    ins &\Coloneqq \exists (r_3 \sqcap r_4)^-.(\exists r_1\nominal{e_1} \sqcap \neg\exists r_2\nominal{e_2}), \\
  \Approximated(ins) &\Coloneqq 
  \begin{aligned}[t]
    & \exists r_3^-.(\exists r_1.\nominal{e_1} \sqcap \neg\exists r_2 \nominal{e_2})\; \sqcap\\
    & \exists r_4^-.(\exists r_1 \nominal{e_1} \sqcap \neg\exists r_2 \nominal{e_2}).
  \end{aligned}
\end{align}

\section{Additional analyses on FB15k-237-DAG-QA and FB15k-DAG-QA datasets}
\label{QueryAns_Analysis}
\begin{figure}[H]
    \centering
    \includegraphics[width=0.8\linewidth]{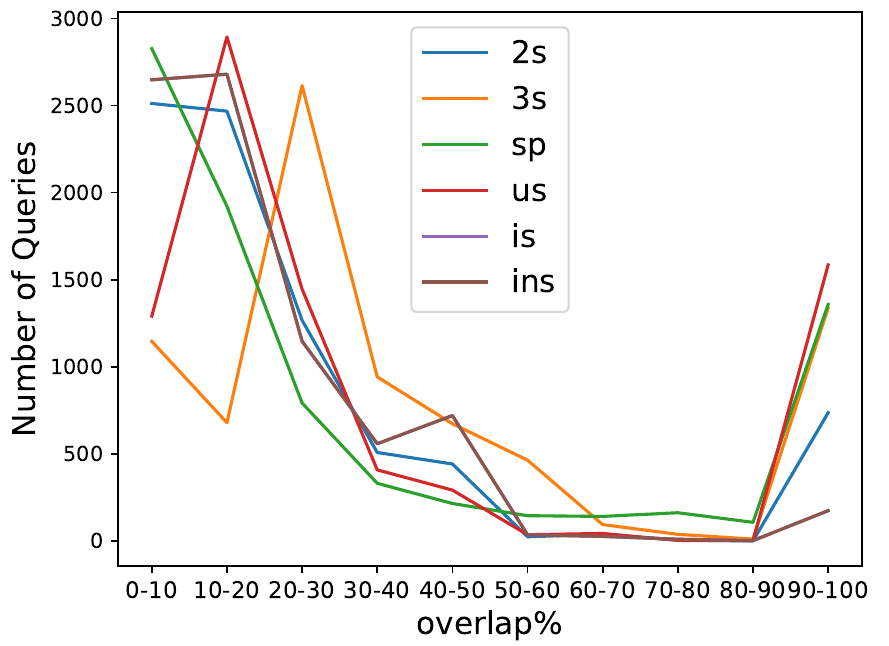}
    \caption{Proportion of overlap between DAG query answers and the corresponding Tree-Form query answers from FB15k-DAG-QA Easy test dataset.}
    \label{FB15kQueryAns_Analysis}
\end{figure}

\begin{figure}[H]
    \centering
    \includegraphics[width=0.8\linewidth]{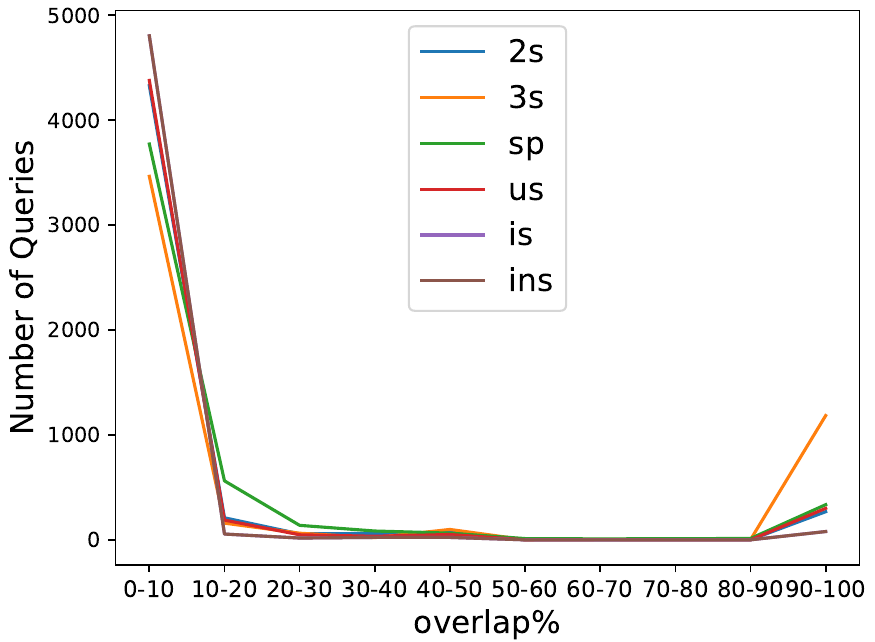}
    \caption{Proportion of overlap between DAG query answers and the corresponding Tree-Form query answers from FB15k-237-DAG-QA Easy test dataset.}
    \label{FB15k-237-QueryAns_Analysis}
\end{figure}

Figures \ref{FB15kQueryAns_Analysis} and \ref{FB15k-237-QueryAns_Analysis} provide additional analysis on the answer sets of the randomly generated DAG queries from FB15k and FB15k-237. 

\section{Further experimental details}
All experiments and the data generation codes are available via \url{https://github.com/RoyaHe/DAG_RC}.
\label{Experimental setup}

\begin{table}[t!]
    \centering
    \caption{Number of train/valid/test queries generated for individual DAG query structure in easy and hard modes.}
    \begin{tabular}{lllll}
    \toprule
    \textbf{Dataset} & \textbf{Train} & \textbf{Valid} & \textbf{Test-Easy} & \textbf{Test-Hard} \\
    \midrule
    NELL-DAG & 10,000 & \hspace{1mm}1000 & \hspace{5mm}1000 & \hspace{5mm}1500\\
    FB15k-237-DAG & 50,000 & \hspace{1mm}1000 & \hspace{5mm}5000 & \hspace{5mm}4700 \\
    FB15k-DAG & 80,000 & \hspace{1mm}8000 & \hspace{5mm}8000 & \hspace{5mm}7500 \\
    \bottomrule
    \end{tabular}
    \label{Num:Test Queries}
\end{table}

\subsection{Hyperparameters and Computational Resource}
All of our experiments are implemented in Pytorch \cite{Pytorch} framework and run on four Nvidia A100 GPU cards. For hyperparameters search, we performed a grid search of learning rates in $\{5\times 10^{-5},10^{-4},5\times10^{-4}\}$, the batch size in $\{256, 512, 1024\}$, the negative sample sizes in $\{128,64\}$, the regularization coefficient $\omega$ in $\{0.02, 0.05, 0.08, 0.1\}$ and the margin $\gamma$ in $\{10,16,24,30,40,60,80\}$. The best hyperparameters are shown in Table \ref{hyperparameters}. 

\begin{table}[H]
  \centering
  \resizebox{\hsize}{!}{
  \begin{tabular}{llllllll}
    \toprule
    \textbf{Dataset} & Model & d & b & n & $\gamma$ & $l$ & $\omega$\\
    \midrule
    & Query2Box ({\OurMethod}) & 400 & 512 & 128 & 24 & 1e-4 & -\\
    NELL-DAG & BetaE ({\OurMethod}) & 400 & 512 & 128 & 60 & 1e-4 & -\\
    & ConE ({\OurMethod}) & 800 & 512 & 128 & 20 & 1e-4 & 0.02\\
    \midrule
    & Query2Box ({\OurMethod}) & 400 & 512 & 128 & 16 & 1e-4 & -\\
    FB15k-237-DAG & BetaE ({\OurMethod}) & 400 & 512 & 128 & 60 & 1e-4 & -\\
    &  ConE ({\OurMethod}) & 800 & 512 & 128 & 30 & 5e-5 & 0.02\\
    \midrule
    & Query2Box ({\OurMethod}) & 400 & 512 & 128 & 16 & 1e-4 & -\\
    FB15k-DAG & BetaE ({\OurMethod}) & 400 & 512 & 128 & 60 & 1e-4 & -\\
    & ConE ({\OurMethod}) & 800 & 512 & 128 & 40 & 5e-5 & 0.02\\
    \bottomrule
  \end{tabular}
  }
  \caption{Hyperparameters found by grid search. d is the embedding dimension, b is the batch size, n is the negative sampling size, $\gamma$ is the margin in loss, l is the learning rate, $\omega$ is the regularization parameter in the distance function.}
  \label{hyperparameters}
\end{table}

\subsection{Further implementation details of {\OurMethod} with additional constraints}
For the regularization of the restricted conjunction preserving tautology, we encourage the tautology $\exists (r \sqcap s). \nominal{e} \equiv \exists r.\nominal{e} \sqcap \exists s.\nominal{e}$ (see Proposition~\ref{prop:tautologies}) with the following loss:
\begin{equation}
    \mathcal{L}_r \Coloneqq \Difference(\embedding{\exists (r \sqcap s). \nominal{e}}, \Intersection(\embedding{r.\nominal{e}}, \embedding{s.\nominal{e}})),
\end{equation}
To enforce the minimization of such loss in our learning objective, we further mine two types of queries from the existing train queries, 2rs and 3rs, that can be expressed as $\ALCOIR$ concepts as follows:
\begin{align}
  2rs &\Coloneqq \exists ((r_1 \sqcap r_2))^-.\nominal{e_1}, \\
  \Approximated(2rs) &\Coloneqq 
  \begin{aligned}[t]
    & \exists (r_1)^-.\nominal{e_1}\; \sqcap\\
    & \exists (r_2)^-.\nominal{e_1},
  \end{aligned} \\
  3rs &\Coloneqq \exists ((r_1 \sqcap r_2 \sqcap r_3)^-.\nominal{e_1}, \\
  \Approximated(3rs) &\Coloneqq
  \begin{aligned}[t]
  &\exists (r_1)^-.\nominal{e_1}\; \sqcap\\
  &\exists (r_2)^-.\nominal{e_1}\; \sqcap\\
  &\exists (r_3)^-.\nominal{e_1}.
  \end{aligned}
\end{align}

\subsection{Computational costs of {\OurMethod}}
To evaluate the training speed, for each model with {\OurMethod}, we calculated the average running time (RT) per 100 training steps on dataset NELL-DAG. For fair comparison with baseline models, we ran all models with the same number of embedding parameters. Integrating {\OurMethod} generally increases the computational cost for existing models. However, models like Query2Box can be enhanced to outperform baseline models like BetaE, while still maintaining lower computational costs of 22s per 100 steps.
\begin{table}[H]
  \caption{Computational costs of {\OurMethod} and the baselines.}
  \label{Computational Costs}
  \centering
  \begin{tabular}{lll}
    \toprule
    \textbf{Model} & \textbf{AVG MRR} & \hspace{5mm} \textbf{RT per 100 steps}\\
    \midrule
    Q2B\cite{querytobox} & \hspace{7mm} 21.23 & \hspace{9mm} 15s\\
    Q2B+{\OurMethod} & \hspace{7mm} \textbf{42.41} & \hspace{9mm} 22s\\
    BetaE\cite{BetaE} & \hspace{7mm} 18.32 & \hspace{9mm} 24s\\
    BetaE+{\OurMethod} & \hspace{7mm} 41.33 & \hspace{9mm} 37s\\
    ConE\cite{ConE} & \hspace{7mm} 26.54 & \hspace{9mm} 18s\\
    ConE+{\OurMethod} & \hspace{7mm} 40.19 & \hspace{9mm} 50s\\
    \bottomrule
  \end{tabular}
\end{table}

\section{Performance of {\OurMethod} on Tree-form queries}
\label{DAGE-Tree-results(complete)}
Table \ref{DAGE-Tree-results} summarizes the performances of query embedding models with {\OurMethod} on existing tree-form query answering benchmark datasets, NELL-QA, FB15k-237-QA and FB15k-QA \cite{BetaE}. Firstly, {\OurMethod} enhances these models by enabling them to handle DAG queries while preserving their original performance on tree-form queries. Secondly, {\OurMethod} shows significant improvement only on DAG queries, with little effect on tree-form queries, supporting our assumption that {\OurMethod} effectively enhances baseline performance for these new query types.

\begin{table*}
    \centering
    \begin{tabular}{llllllllllll}
    \toprule
         \textbf{Dataset} & \textbf{Model} & \hspace{2mm}\textbf{1p} & \hspace{2mm}\textbf{2p} & \hspace{2mm}\textbf{3p} & \hspace{2mm}\textbf{2i} & \hspace{2mm}\textbf{3i} & \hspace{2mm}\textbf{pi} & \hspace{2mm}\textbf{ip} & \hspace{2mm}\textbf{2u} & \hspace{2mm}\textbf{up} & \textbf{AVG}\\
    \midrule
          & Q2B &  42.7 & 14.5 & 11.7 & 34.7 & 45.8 & 23.2 & 17.4 & 12.0 & 10.7 & 23.6\\
          
          & Q2B ({\OurMethod}) &  42.09  & 23.39  & 21.28  & 28.64 & 41.09 & 20.0 & 12.30 & 27.51 & 15.86 & 28.3 \\
          
          & BetaE & 53.0 & 13.0 & 11.4 & 37.6 & 47.5 & 24.1 & 14.3 & 12.2 & 8.5 & 24.6\\

          NELL-QA & BetaE ({\OurMethod}) & 53.4 & 12.9 & 10.8 & 37.6 & 47.1 & 23.8 & 13.8 & 12.3 & 8.3 & 24.4\\
          
           & ConE &  53.1 & 16.1 & 13.9 & 40.0 & 50.8 & 26.3 & 17.5 & 15.3 & 11.3 & 27.2 \\

          & ConE ({\OurMethod}) & 53.2 & 15.7 & 13.7 & 39.9 & 50.7 & 26.0 & 17.0 & 14.8 & 10.9 &  26.8 \\
          

    \midrule
          & Q2B & 41.3 & 9.9 & 7.2 & 31.1 & 45.4 & 21.9 & 13.3 & 11.9 & 8.1 & 21.1\\

          & Q2B ({\OurMethod}) & 42.6 & 11.4 & 9.3 & 30.2 & 42.8 &  22.4 & 12.1 & 12.1 & 9.2 & 21.4\\
          
          & BetaE & 39.0 & 10.9 & 10.0 & 28.8 & 42.5 & 22.4 & 12.6 & 12.4 & 9.7 & 20.9\\

          FB15k-237-QA & BetaE ({\OurMethod}) & 38.9 & 10.87 & 9.94 & 29.1 & 42.7 & 22.0 & 11.0 & 12.1 & 9.6 & 20.7\\
          
           & ConE & 41.8 & 12.8 & 11.0 & 32.6 & 47.3 & 25.5 & 14.0 & 14.5 & 10.8 & 23.4\\

          & ConE ({\OurMethod}) & 42.13 & 12.8 & 10.8 & 32.6 &  47.0 & 25.4 & 13.2  & 14.3  & 10.5 & 23.2\\
          

    \midrule
          & Q2B & 70.5 & 23.0 & 15.1 & 61.2 & 71.8 & 41.8 & 28.7 & 37.7 & 19.0 & 40.1\\

          & Q2B ({\OurMethod}) & 67.9 & 24.5 & 21.3 & 53.4 & 64.82 &  40.3 & 23.5  & 35.2  &  21.7 & 39.2\\

          & BetaE & 65.1 & 25.7 & 24.7 & 55.8 & 66.5 & 43.9 & 28.1 & 40.1 & 25.2 & 41.6\\

          FB15k-QA & BetaE ({\OurMethod}) & 64.5 & 24.6 & 23.6 & 55.6 & 66.5  & 42.8 & 22.5 & 40.2 & 23.8 & 40.5\\ 
          
           & ConE & 73.3 & 33.8 & 29.2 & 64.4 & 73.7 & 50.9 & 35.7 & 55.7 & 31.4 & 49.8\\

          & ConE ({\OurMethod}) &  74.3 & 31.9 & 27.4  &  63.9 & 73.5  &  50.1 & 29.8  & 53.6 & 29.4 & 48.2\\
          

    \bottomrule
    \end{tabular}
    \caption{MRR performance of the retrained baseline models with {\OurMethod} method on non-neg tree-form query benchmark datasets}
    \label{DAGE-Tree-results}
\end{table*}

\begin{table*}
    \centering
    \begin{tabular}{llllllll}
    \toprule
         \textbf{Dataset} & \textbf{Model} & \hspace{2mm}\textbf{2in} & \hspace{2mm}\textbf{3in} & \hspace{2mm}\textbf{inp} & \hspace{2mm}\textbf{pin} & \hspace{2mm}\textbf{pni} & \hspace{2mm}\textbf{AVG}\\
    \midrule
          & BetaE & \hspace{2mm} 5.1 & \hspace{2mm} 7.8 & \hspace{2mm} 10.0 & \hspace{2mm} 3.1 & \hspace{2mm} 3.5 &  \hspace{2mm} 5.9\\

          NELL-QA & BetaE ({\OurMethod}) & \hspace{2mm} 5.3 &  \hspace{2mm} 7.9 &  \hspace{2mm}10.1 &  \hspace{2mm}  2.9&  \hspace{2mm} 3.7 &  \hspace{2mm} 5.98\\
          
           & ConE &  \hspace{2mm} 5.7 & \hspace{2mm} 8.1 & \hspace{2mm} 10.8 & \hspace{2mm} 3.5 & \hspace{2mm} 3.9 &  \hspace{2mm} 6.4\\

          & ConE ({\OurMethod}) &  \hspace{2mm} 5.5&  \hspace{2mm} 8.0 &  \hspace{2mm} 10.9&  \hspace{2mm} 3.8&  \hspace{2mm} 4.0 &  \hspace{2mm} 6.44 \\
          

    \midrule
          & BetaE &  \hspace{2mm} 5.1 &  \hspace{2mm} 7.9 &  \hspace{2mm} 7.4 &  \hspace{2mm} 3.6 &  \hspace{2mm} 3.4 &  \hspace{2mm} 5.4\\
          FB15k-237-QA & BetaE ({\OurMethod}) &  \hspace{2mm} 5.2 &  \hspace{2mm} 8.1 &  \hspace{2mm} 7.1&  \hspace{2mm} 3.5&  \hspace{2mm} 3.6 &  \hspace{2mm}  5.5\\
          
           & ConE &  \hspace{2mm} 5.4 &  \hspace{2mm} 8.6 &  \hspace{2mm} 7.8 &  \hspace{2mm} 4.0 &  \hspace{2mm} 3.6 &  \hspace{2mm} 5.9\\
          & ConE ({\OurMethod}) &  \hspace{2mm} 5.1 &  \hspace{2mm} 8.7 &  \hspace{2mm} 7.9 &  \hspace{2mm} 3.8 &  \hspace{2mm} 3.4 &  \hspace{2mm} 5.78 \\
          

    \midrule
          & BetaE &  \hspace{2mm} 14.3 &  \hspace{2mm} 14.7 &  \hspace{2mm} 11.5 &  \hspace{2mm} 6.5 &  \hspace{2mm} 12.4 &  \hspace{2mm} 11.8\\
          FB15k-QA & BetaE ({\OurMethod}) &  \hspace{2mm} 14.1 &  \hspace{2mm} 14.5 &  \hspace{2mm} 11.4 &  \hspace{2mm} 6.4 &  \hspace{2mm} 12.6 &  \hspace{2mm} 11.8\\
          
           & ConE &  \hspace{2mm} 17.9 &  \hspace{2mm} 18.7 &  \hspace{2mm} 12.5 &  \hspace{2mm} 9.8 &  \hspace{2mm} 15.1 &  \hspace{2mm} 14.8\\
          & ConE ({\OurMethod}) &  \hspace{2mm} 17.8 &  \hspace{2mm} 18.9 &  \hspace{2mm} 12.1 &  \hspace{2mm} 10.1 &  \hspace{2mm} 15.3 &  \hspace{2mm} 14.84\\

    \bottomrule
    \end{tabular}
    \caption{MRR performance of the retrained baseline models with {\OurMethod} method on tree-form queries with negation on benchmark datasets}
\end{table*}

\section{Comparison with other query embedding methods}
\label{Comparison with other query embedding methods}
To further assess the effectiveness of {\OurMethod}, we compare the baseline models enhanced by {\OurMethod} with two prominent query embedding models, CQD \cite{CQD} and BiQE \cite{BiQE}. Both models are theoretically believed to be capable of handling DAG queries based on their design. Table \ref{easy_results_withothers} and \ref{hard_results_withothers} summarize the performances of these models on our proposed DAG queries benchmark datasets. These methods enhanced with {\OurMethod} consistently outperform CQD and BiQE across all types of DAG queries and datasets. Although some baseline methods perform significantly worse than BiQE and CQD in tree-form query answering tasks (according to the reported results from BiQE and CQD), their integration with {\OurMethod} allows them to improve and surpass those models on the new DAG query benchmark datasets. This further supports our argument regarding the effectiveness of {\OurMethod}.
 
\begin{table*}[h]
    \centering
    \begin{tabular}{llllllllll}
    \toprule
         \textbf{Dataset} & \textbf{Model} & \hspace{2mm}\textbf{2s} & \hspace{2mm}\textbf{3s} & \hspace{2mm}\textbf{sp} & \hspace{2mm}\textbf{is} & \hspace{2mm}\textbf{us} & \hspace{2mm}\textbf{Avg}$_{nn}$ & \hspace{2mm}\textbf{ins} & \hspace{2mm}\textbf{Avg}\\
    \midrule
          & Query2Box ({\OurMethod}) & \textbf{37.61}   & 49.42   & \textbf{41.71}   & \textbf{40.57}   & \textbf{42.75}   & \hspace{2mm}\textbf{42.41}   & \hspace{3mm}- & \hspace{3mm}-\\

          & BetaE ({\OurMethod}) & 36.87   & 57.14   & 34.95   & 39.90  &  37.80  & \hspace{1mm} 41.33   & \textbf{34.68}   & \textbf{40.22}  \\

          NELL-DAG & ConE ({\OurMethod}) & 33.50   & \textbf{57.35}   & 38.43   & 37.93   & 33.74   & \hspace{1mm} 40.19   & 33.94   & 39.15   \\
          
          & CQD-Beam & 22.60&  44.84&  17.51&  24.88&  1.60 & \hspace{1mm} 22.29& \hspace{3mm}- & \hspace{3mm}-\\

          & BiQE & 20.74 &  45.38 &  20.76 &  28.37 &  \hspace{1mm} - & \hspace{1mm} 28.81 & \hspace{3mm}- & \hspace{3mm}-\\
    \midrule
          & Query2Box ({\OurMethod}) & \textbf{7.41}   & \textbf{12.64}   & 10.07   & \textbf{7.32}   & \textbf{5.03}   & \hspace{2mm}8.49   & \hspace{3mm}- & \hspace{3mm}-\\

          & BetaE ({\OurMethod}) & 6.27   & 12.11   & 9.64   & 6.66   &  4.09   & \hspace{2mm}7.75   & \textbf{6.58}   & 7.56   \\

          FB15k-237-DAG & ConE ({\OurMethod}) & 6.87  & 11.66  & \textbf{12.36}  & 6.90   & 4.80  & \hspace{2mm}\textbf{9.54}  & 6.08   & \textbf{8.12}  \\
          
          & CQD-Beam & 4.31 & 8.65 & 5.56 & 3.71 & 0.13 & \hspace{2mm}4.47 & \hspace{3mm}- & \hspace{3mm}-\\

          & BiQE & 2.11 & 2.74 & 4.08 & 5.73 & \hspace{1mm} - & \hspace{2mm}3.67 & \hspace{3mm}- & \hspace{3mm}-\\

    \midrule

          & Query2Box ({\OurMethod}) & 37.74   & 42.93   & 24.30   & 29.37   & 25.97   & \hspace{1mm} 31.46   & \hspace{3mm}- & \hspace{3mm}- \\

          & BetaE ({\OurMethod}) & 32.65    & 46.17    & 32.48   & 28.15    & 28.10   & \hspace{1mm} 33.50   & 25.39   & 32.15   \\

          FB15k-DAG & ConE ({\OurMethod}) & \textbf{41.67}   & \textbf{56.70}   & \textbf{33.36}   & \textbf{36.54}  & \textbf{32.36}   & \hspace{1mm} \textbf{40.12}   & \textbf{30.86}   & \textbf{38.58}  \\

          & CQD-Beam & 22.21 & 36.77 & 23.44 & 15.18 & 1.64 & \hspace{1mm} 19.85 & \hspace{3mm}- & \hspace{3mm}-\\

          & BiQE & 26.31 & 35.12. & 20.37 & 20.08 & \hspace{1mm} - & \hspace{1mm} 22.25 & \hspace{3mm}- & \hspace{3mm}-\\
    \bottomrule
    \end{tabular}
    \caption{The table presents the MRR performance of baseline models integrated with {\OurMethod} on easy benchmark datasets, in comparison with CQD\cite{CQD} and BiQE\cite{BiQE}.}
    \label{easy_results_withothers}
\end{table*}

\begin{table*}[h]
    \centering
    \begin{tabular}{llllllllll}
    \toprule
         \textbf{Dataset} & \textbf{Model} & \hspace{2mm}\textbf{2s} & \hspace{2mm}\textbf{3s} & \hspace{2mm}\textbf{sp} & \hspace{2mm}\textbf{is} & \hspace{2mm}\textbf{us} & \hspace{2mm}\textbf{Avg}$_{nn}$ & \hspace{2mm}\textbf{ins} & \hspace{2mm}\textbf{Avg}\\
    \midrule
          & Query2Box ({\OurMethod}) &  25.38   & 20.13   & 21.25   & 24.85   & 29.24   & \hspace{1mm} 24.17   & \hspace{3mm}- & \hspace{3mm}-\\

           & BetaE ({\OurMethod}) & 27.68   & 32.25   & 16.36   & 26.14   & 29.19   & \hspace{1mm} 26.32  & 33.64   & 27.54  \\

          NELL-DAG & ConE ({\OurMethod}) &  \textbf{30.71}   & \textbf{38.41}   & \textbf{24.76}   & \textbf{28.44}   & \textbf{31.06}   & \hspace{1mm} \textbf{30.67}   & \textbf{34.07} & \textbf{31.24}   \\

          & CQD-Beam & 12.25 & 21.73 & 10.78 & 12.73 & 2.17& \hspace{1mm} 11.93 & \hspace{3mm}- & \hspace{3mm}-\\

          & BiQE & 13.57 & 19.46 & 12.03 & 15.38 & \hspace{3mm}- & \hspace{1mm} 15.11 & \hspace{3mm}- & \hspace{3mm}-\\

    \midrule
          & Query2Box ({\OurMethod}) & 4.81   & \textbf{2.81}   & 7.87   & \textbf{5.26}   & \textbf{4.38}   & \hspace{1mm}\textbf{6.95}   & \hspace{3mm}- & \hspace{3mm}-\\
          
           & BetaE ({\OurMethod}) & \textbf{4.89}   & 1.66   & 8.28   & 4.75   & 3.50   & \hspace{1mm}4.61   & \textbf{6.06}   & 4.85 \\

          FB15k-237-DAG & ConE ({\OurMethod}) & 4.78   & 2.09 & \textbf{9.72}   & 4.84    & 4.16    & \hspace{1mm}5.12   & 5.25   & \textbf{5.14}  \\
          
          & CQD-Beam & 2.74 & 1.63 & 4.63 & 2.38 & 0.10 & \hspace{1mm}2.29 & \hspace{3mm}- & \hspace{3mm}- \\

          & BiQE & 3.13 & 2.01 & 3.37 & 1.89 & \hspace{3mm}- & \hspace{1mm} 2.60& \hspace{3mm}- & \hspace{3mm}- \\
    \midrule
          & Query2Box ({\OurMethod}) & 33.78   & 39.67   & 19.61   & 26.91   & 24.76   & 28.95   & \hspace{3mm}- & \hspace{3mm}- \\

          & BetaE ({\OurMethod}) & 30.57   & 44.30   & 29.35   & 25.72   & 26.63   & 31.31   & 25.18   & 30.29  \\

         FB15k-DAG & ConE ({\OurMethod}) & \textbf{40.14}   & \textbf{57.06}   & 29.23 & \textbf{34.63}   & \textbf{31.45}   & \textbf{38.50}   & \textbf{30.74}   & \textbf{37.21}  \\

          & CQD-Beam & 19.90 & 33.88 & 22.43 & 12.66 & 1.60 & 18.09 & \hspace{3mm}- & \hspace{3mm}-\\

          & BiQE & 17.36 & 30.37 & 25.38 & 15.89 & \hspace{3mm}- & 22.25 & \hspace{3mm}- & \hspace{3mm}-\\
    \bottomrule
    \end{tabular}
    \caption{The table presents the MRR performance of baseline models integrated with {\OurMethod} on hard benchmark datasets, in comparison with CQD\cite{CQD} and BiQE\cite{BiQE}.}
    \label{hard_results_withothers}
\end{table*}

\section{Performances of DAGE with additional logical constraints}
Table \ref{Complete Results: DAGE with logical constraints} summarizes the performances of all baseline methods enhanced by DAGE with additional logical constraints, i.e., monoticity and restricted conjunction preservation in proposition \ref{prop:tautologies}.
\label{Performances of DAGE with additional logical constraints}
\begin{table*}
    \centering
    \renewcommand{\arraystretch}{0.85} 
     \caption{The MRR performance of baseline methods with {\OurMethod} on NELL-DAG hard benchmark dataset, along with their performance when integrated with additional logical constraints.}
       \vspace{-0.2cm}
    \label{Complete Results: DAGE with logical constraints}
    \begin{tabular}{llllllllll}
    \toprule
          \textbf{Model} & \hspace{2mm}\textbf{2s} & \hspace{2mm}\textbf{3s} & \hspace{2mm}\textbf{sp} & \hspace{2mm}\textbf{is} & \hspace{2mm}\textbf{us} & \hspace{2mm}\textbf{Avg}$_{nn}$ & \hspace{2mm}\textbf{ins} & \hspace{2mm}\textbf{Avg}\\
          \midrule

           Query2Box ({\OurMethod}) & 25.38 & 20.13  & 21.25 & 24.85 & 29.24 & \hspace{1mm} 24.17 & \hspace{3mm}- & \hspace{3mm}-\\

           Query2Box ({\OurMethod}+Distr) &  \textbf{25.93} & 21.50	& 20.85	& 24.73	& 29.71 & \hspace{1mm} 24.54 & \hspace{3mm}- & \hspace{3mm}-\\

           Query2Box ({\OurMethod}+Mono) &  25.90 & 21.74 & 21.87 &	25.41 & \textbf{30.17} & \hspace{1mm} 25.02 & \hspace{3mm}- & \hspace{3mm}-\\

           Query2Box ({\OurMethod}+Distr+Mono) &25.87&\textbf{22.01}&\textbf{22.34}& \textbf{24.96}&30.02& \hspace{1mm} \textbf{25.04} & \hspace{3mm}- & \hspace{3mm}-\\
          
          \midrule

           BetaE ({\OurMethod}) & 27.68 & 32.25 & 16.36 & 26.14 & 29.19 & \hspace{1mm} 26.32 & 33.64 & 27.54 \\

           BetaE ({\OurMethod}+Distr) & 27.91 &32.87&17.12&\textbf{27.02}&\textbf{30.13} &\hspace{1mm} 27.01&34.28&28.22\\

           BetaE ({\OurMethod}+Mono) & 28.01 & \textbf{33.56} & 16.89 & 26.93  & 29.47 &\hspace{1mm} 26.97 & 34.17 & 28.17\\

           BetaE ({\OurMethod}+Distr+Mono) & \textbf{28.11}	&  33.48&\textbf{17.67}&	26.83&29.36&\hspace{1mm} \textbf{27.09}&\textbf{34.49}&\textbf{28.32}\\
          
          \midrule
           ConE ({\OurMethod}) &  30.71 & 38.41 & 24.76 & 28.44 & 31.06 & \hspace{1mm} 30.67 & 34.07 & 31.24 \\

           ConE ({\OurMethod}+Distr) & 31.23 & 39.37 & 25.08 & \textbf{28.73} & \textbf{31.92}&\hspace{1mm} 31.27 & \textbf{36.34} & 32.11 \\

           ConE ({\OurMethod}+Mono) &  31.47&39.82&25.17&28.57&31.24&
           \hspace{1mm} 31.25&35.84&32.02 \\

           ConE ({\OurMethod}+Distr+Mono) & \textbf{31.88}&\textbf{39.89}&\textbf{25.28}&28.64&31.57 & \hspace{1mm} \textbf{31.45} &35.93 & \textbf{32.20} \\
          

    \bottomrule
    \end{tabular}
\end{table*}

\end{document}